\documentclass[12pt]{iopart}
\usepackage{iopams}  
\usepackage[utf8]{inputenc}
\usepackage{graphicx, subfigure}
\usepackage[colorlinks,allcolors=cyan!70!black]{hyperref}
\usepackage{color}
\expandafter\let\csname equation*\endcsname\relax

\expandafter\let\csname endequation*\endcsname\relax
\usepackage{amsmath}
\usepackage{amsfonts}
\usepackage{amssymb}
\usepackage{graphicx, subfigure}
\usepackage[dvipsnames]{xcolor}

\usepackage{tikz}
\usetikzlibrary{arrows,decorations.pathmorphing,backgrounds,positioning,fit,petri}
\usetikzlibrary{shapes.geometric}
\usepgflibrary{shapes.symbols}
\usetikzlibrary{circuits.ee.IEC}
\usetikzlibrary{patterns} 
\usetikzlibrary{backgrounds}
\DeclareMathOperator{\el}{\overline{e}}
\usepackage [basic,optics, gate,box]{circ}
\begin{document}

\title[Bio-Impedance Spectroscopy: Basics and Applications]{Bio-Impedance Spectroscopy: Basics and Applications}

\author{Daniil D. Stupin$^1$, Ekaterina A. Kuzina$^1$, Anna A. Abelit$^{1,2}$, Sergei V. Koniakhin$^{1,3}$, Anton E. Emelyanov$^{1,4}$, Dmitrii~M.~Nikolaev$^{1}$, Mikhail N. Ryazantsev$^{1,5}$, and Michael~V.~Dubina$^{6}$}

\address{$^1$Alferov University, 8/3 Khlopina street, Saint Petersburg 194021, Russia\\
$^2$Peter the Great St.Petersburg Polytechnic University, 195251, Russia, St.Petersburg, Polytechnicheskaya, 29\\
$^3$Institut Pascal, PHOTON-N2, Universit\'e Clermont Auvergne, CNRS, SIGMA Clermont, F-63000 Clermont-Ferrand, France\\
$^4$Pavlov First Saint Petersburg State Medical University, L'va Tolstogo str. 6-8, Saint Petersburg, Russia
197022\\
$^5$Institute of Chemistry, Saint Petersburg State University, 26 Universitetskii pr, Saint Petersburg, Russia 198504
\\$^6$Institute of Highly Pure Biopreparation of the Federal Medical-Biological Agency, Pudozhskaya 7, St. Petersburg, 197110\\

}
\ead{Stu87@ya.ru, Stupin@spbau.ru}

\vspace{10pt}
\begin{indented}
\item[]January 2020
\end{indented}

\begin{abstract}
In this review, we aim to introduce the reader to the technique of electrical impedance spectroscopy (EIS) with a focus on its biological and medical applications. We explain the theoretical and experimental aspects of the EIS with the details essential for biological studies, \textit{i.e.} interaction of metal electrodes with biological matter and liquids, strategies of increasing measurement rate and noise reduction in bio-EIS experiments \textit{etc}. We give various examples of successful bio-EIS practical implementations in science and technology: from the whole-body health monitoring and sensors for vision prosthetic care to single living cell examination platforms and virus diseases research. Present review can be used as a bio-EIS tutorial for students as well as a handbook for scientists and engineers due to extensive references covering the contemporary research papers in the field.

\end{abstract}

\section{Introduction}

Since 1886, when electrical impedance spectroscopy (EIS) was originally conceptualized by Oliver Heaviside \cite{heaviside2011electrical,macdonald2006reflections}, it has evolved into a powerful and widely used experimental technique \cite{Barsoukov,Lvovich,Review,BioImp,kanoun2018impedance}.
Despite the simple concept -- measuring the electrical current passing through the sample at various frequencies of the excitation voltage -- this technique is currently applied in numerous domains.
For example, EIS is commonly used as a testing technique for modern electronic devices: from semiconductor hetero-structures \cite{brus2012impedance} to audiophile amplifiers \cite{nielsen2003active};
EIS is a state-of-the-art method in electrochemistry and materials science: from nano-objects investigation and alternative energy sources characterization to Mars surface research \cite{barbero2007evidence,lelidis2005effect, talaga2013electrochemical,SolarCell,Wang,trautner2003detection}. Being a label-free non-optical non-destructive and easy to implement technique, EIS became a promising experimental approach in biological and medical applications, such as biosensing technologies \cite{BioImp,Geaever,Giaever_book,Cell_based_biosensors, Dittami,cheran2008probing} and diagnosis of diseases, including cancer and virus detection \cite{ICG,zou2003review,brown2001medical,nidzworski2014universal,diouani2008miniaturized,hassen2011quantitation,campbell2007monitoring,mccoy2005use,pennington2017electric,cho2007impedance,golke2012xcelligence,teng2013real}.

Historically, one of the first biological EIS applications was proposed in 1926 by Leon~S.~Theremin, who developed a contactless musical instrument -- \textit{thereminvox} -- that converts the impedance of the performer's hands into various music tones \cite{Rzevkin1948_Eng, theremin1996design}.
Almost at the same time, in~1928, Kenneth S. Cole developed the theoretical basis for the description of the tissue impedance, and later in 1937, together with Howard~J.~Curtis he performed one of the first EIS investigations of large living cells, namely the giant plant cells  \textit{Nitella} \cite{cole_1,cole_2,curtis1937transverse}.
In 1992, Charles R. Keese and Ivar Giaever take advantage of the remarkable progress in the microelectronics fabrication to extend EIS applications for studying  a wide range of living cells, including mammalian cells, and bring into life this idea in \textit{electrical cell-substrate impedance sensing} (ECIS) technology \cite{Geaever,Giaever_book,tiruppathi1992electrical, Giaever2}, which for now days became an outstanding tool for label-free and real-time cells research.
ECIS device uses the modified Petri dish with the planar electrodes located on it's bottom on which investigated living cells are seeded.
As far as the cells affect significantly the electrodes impedance \cite{Geaever} it becomes possible to determine their properties by studying ECIS impedance and its evolution.
For example, the ECIS methodology allows measuring the number of cells adherent to the electrodes \cite{xiao2003line,szulcek2014electric}, estimating the viability of cells \cite{stupin2018single,gonzalez2018electrical}, studying cell motility \cite{szulcek2014electric}, investigating cell-cell interactions \cite{tran2016electric}, quantifying the activity of pharmaceuticals \cite{Giaever_book,asphahani2007cellular,keese2004electrical,asphahani2011single}, \textit{etc}.

This extensive EIS utilization in the science resulted in appearance of the large number of remarkable text-books, monographs, and review papers devoted to the impedance spectroscopy \cite{Barsoukov,Lvovich,Review,BioImp,kanoun2018impedance,C_Spectroscopy,Giaever_book}. Moreover, a significant part of information on EIS grounds can be found in stand-alone chapters in almost every physics and electronics textbooks \cite{Feynman_Lectures,Tietze}.
However, most number of these texts either focus mainly on applications, not explaining enough the underlying theory, or focus on the detailed explanation of the EIS theory, making it difficult for the first reading.
Thus in this tutorial review  we make effort to keep the balance for EIS explanation.
To be specific, we try to make introduction to the theory, experiment, and bio-applications of the EIS from the first principles. We explain the plans of the problems solution and stress the most important points. As a consequence, to read this this tutorial only the basic knowledge in differential and integral calculus, complex variable theory, and general physics are required. This paper can be used by students as first-reading reference in the bio-EIS area.

The article is organized as follows.
First, we will introduce the theoretical basis for EIS with a focus on the details that are important for biological applications.
Then, we will consider the experimental aspects of EIS, and also compare different impedance-measurement techniques.
Finally, we will discuss the application of EIS in biology and medicine with paying special attention to cell research, health monitoring, and decease diagnostics such as viral infection and cancer.\\

\section{Theoretical aspects of EIS}
\subsection{Electrical Immittance}
\textit{Impedance} is complex-valued electrical resistance that depends on the frequency $\omega$ of excitation voltage (EV).
Typically, impedance is denoted as $Z$, or $Z(\omega)$ stressing the frequency dependence. A closely related quantity is \textit{admittance}, or complex-valued conductivity, that is defined as $\dfrac{1}{Z}$ and typically denoted as $Y$.
From the complex quantities of impedance and admittance, one can derive a number of meaningful characteristics, including real parts (denoted as $\Re$ or by placing apostrophe $^\backprime$), imaginary parts (so-called \textit{reactances}, denoted as $X$, $\Im$, or by placing double apostrophe $^{\backprime\backprime}$), phases, magnitudes, \textit{etc}.
For convenience, to refer to impedance, admittance or any physical quantity that is a function of impedance or admittance, a generic term \textit{immittance} is often used. 

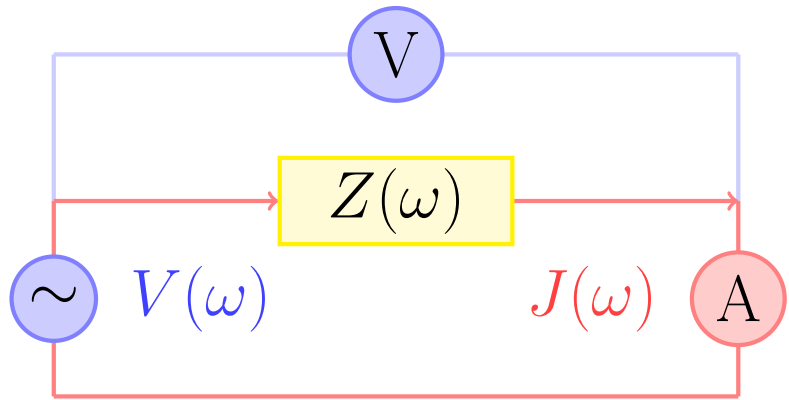
\begin{figure}[h!]
    \centering
\centering
\begin{tikzpicture}[ultra thick, scale=1.3]
\draw [ color=blue!20]   (-3.5,0) -- (-3.5,1.5) ;
\draw [ color=blue!20] (-3.5,1.5) -- (3.5,1.5) ;
\draw  [ color=blue!20](3.5,1.5) -- (3.5,0) ;
\draw [->, color=red!50]  (-3.5,0) -- (-1.2,0) ;
\draw [->, color=red!50]  (0.9,0) -- (3.5,0) ;
\draw [ color=red!50]  (3.5,0) -- (3.5,-2) ;
\draw [ color=red!50]  (3.5,-2) -- (-3.5,-2) ;
\draw [ color=red!50]  (-3.5,-2) -- (-3.5,0) ;
\node at (0,1.5) [circle,draw=blue!50,fill=blue!20] {\huge V};
\node at (0,0) [rectangle,draw=yellow!100,fill=yellow!20] {\huge \phantom{Z}$Z(\omega)$\phantom{Z}};
\node at (-3.5,-1) [circle,draw=blue!50,fill=blue!20] {\huge $\sim$};
\node at (3.5,-1) [circle,draw=red!50,fill=red!20] {\huge A};
\node at (-2,-1) [rectangle] {\textcolor{blue!75}{\huge $V(\omega)$}};
\node at (2,-1) [rectangle] {\textcolor{red!75}{\huge $J(\omega)$}};
\end{tikzpicture} 
    \caption{The common immittance measurement scheme. The excitation voltage $V(\omega)$ is applied to the sample with impedance $Z(\omega)$, which results in current response $J(\omega)$ through it. Excitation voltage and current response are measured with the voltmeter V and the ammeter A, respectively, to derive impedance $Z(\omega)$. \label{Fig:IS_measurement}}
  
\end{figure}

For the single-frequency sine-shape excitation voltage (EV)
\begin{equation}
\label{Eq:Single-Sine_Voltage}
V(t)=V_0 e^{i \omega t}    
\end{equation} 
the immittance could be measured as shown in Fig. \ref{Fig:IS_measurement}. 
If a sample is linear (\textit{i.e.} the superposition principle for excitation voltages and corresponding current responses is fulfilled), the current response will be also a sine-shaped function with frequency $\omega$

\begin{equation}
 J(t)=J_0 e^{i \omega t+i\varphi},   
\end{equation}
where $J_0=J_0(\omega)$ is the magnitude of the current, $\varphi=\varphi(\omega)$ is the current phase-shift with respect to the applied voltage. 
So, according to Ohm's law, the impedance at frequency $\omega$ can be defined as 

\begin{equation}
\label{Eq:immittance}
 Z(\omega)=\frac{1}{Y(\omega)}=\frac{V(t)}{J(t)}=\frac{V_0}{J_0}e^{-i\varphi}=|Z|e^{i\phi}=Z^\backprime+i Z^{\backprime\backprime},
\end{equation}
where $|Z|=V_0/J_0$ is magnitude of the impedance, $\phi=-\varphi$ is phase of the impedance, $Z^\backprime=|Z| \cos \phi$  is real part of the impedance, $Z^{\backprime\backprime}=|Z| \sin \phi$ is imaginary part of the impedance.
Remarkably, that expression \eqref{Eq:immittance} is \textit{time-independent.} 

Now by using superposition principle and  equation \eqref{Eq:immittance}, we can easily calculate the current response for a non-sine, arbitrary-shaped excitation voltage.
Based on the fact that any physically realistic EV can be represented as sum of the single-frequency harmonics of the form Eq. \eqref{Eq:Single-Sine_Voltage} or in more general form as a Fourier integral 
\begin{equation}
V(t)=\int\limits^\infty_{-\infty} \tilde{V}(\omega) e^{i\omega t} d \omega,
\end{equation}
one can start with calculating the response to each harmonic of the excitation voltage using the equation \eqref{Eq:immittance}. Here the amplitude $V_0$ from Eq. \eqref{Eq:Single-Sine_Voltage} is naturally   replaced by $\tilde{V}(\omega)$ -- a complex amplitude of the harmonic with frequency $\omega$.
At the next step, one can calculate the overall current response as the sum (or integral) of current responses at each frequency
\begin{equation}
\label{Eq:Current}
J(t)=\int\limits^\infty_{-\infty} \frac{\tilde{V}(\omega)}{Z(\omega)} e^{i\omega t} d \omega=\int\limits^\infty_{-\infty} Y(\omega)\tilde{V}(\omega) e^{i\omega t} d \omega.
\end{equation}
From Eq. \eqref{Eq:Current} one can see that the immittance completely defines the electrical current through the sample for an arbitrary excitation voltage, or in other words, the immittance contains full information about sample electrical properties.
That's why the immittance concept is very useful for electrical characterisation. 

Moreover, despite that definition of the immittance in Eq. \eqref{Eq:immittance} is applicable only to linear samples (the sample does not produce the overtones), it's is possible to extend EIS application for non-linear systems too. $E.g.$, one can use the small-signal approximation, which is valid if the amplitude of an excitation voltage is small enough to eliminate sample's non-linearity.
Typically, in this type of experiments, a large constant offset voltage is added  to a small-signal alternating EV to obtain additional experimental data \cite{C_Spectroscopy}. Particularly, this technique is a state-off-the-art for semiconductors diagnostics.

The immittance of any real system has several common properties, which we will briefly discussed here without giving the proofs, which can be found in Ref. \cite{Barsoukov, Lvovich}.

\begin{enumerate}
    \item The real and the imaginary parts of the immittance are related to each other via Kramers-Kronig relations \cite{hu1989kramers}.
    \item The real part and the magnitude of immittance are even functions of frequency.
    \item The imaginary part and the phase of the immittance are odd functions.
    \item The poles of the immittance have a negative real part.
\end{enumerate}

\subsection{The concept of equivalent circuits}
One of the main advantages of EIS as an experimental technique is the possibility to analyze the obtained immittance spectra by means of the equivalent circuits approach \cite{Barsoukov,Lvovich}. 
The key idea of this method consist in constructing an equivalent electrical circuit that has the same immittance as an investigated sample.
For example, a suitable equivalent circuit for an ideal semiconductor diode is a serial RC-circuit, because the pn-junction has zero conductivity at constant current and high conductivity at high-frequency current \cite{C_Spectroscopy} (here we assume that EV has small amplitude and rectification effects are eliminated).
As a result, we can associate the capacitance in equivalent circuit with a space charge region of pn-junction, and the resistance in the equivalent circuit with the bulk resistance of the diode base.
Alternatively, for a living cell membrane, a suitable equivalent circuit is the parallel RC-circuit, because it has low (but non-zero) conductivity at constant current and very high conductivity at high-frequency current.
Here, the capacitor corresponds to the geometrical capacitance of the cell membrane, and the resistor corresponds to the resistance of the membrane. 
Thus, the equivalent circuit approach allows direct associating the immittance spectrum with physical and chemical phenomena in the investigated sample and extracting its characteristics.

To determine the topology of the equivalent circuit of the sample, it is often useful to perform a preliminary analysis by plotting special graphs -- Bode diagram (immittance magnitude and phase \textit{v.s.} frequency) and Nyquist plot that is also called the immittance locus (the imaginary part of immittance \textit{v.s.} real part of the immittance). Nyquist plot is in fact a paramteric plot of these two function with the parameter being a frequency.
As an example for this procedure, let's consider a serial RC circuit (see Table~1). Its impedance and admittance are described as

\begin{equation}
    Z=R+\frac{1}{i \omega C}=\sqrt{R^2+\frac{1}{(\omega C)^2}}\times e^{-i\arctan[ 1/(\omega RC)]},
\end{equation}
\begin{equation}
    Y=\frac{1}{Z}=\frac{\omega^2 R C^2}{1+(\omega RC)^2}+i\frac{\omega C}{1+(\omega RC)^2},
\end{equation}
where $R$ is the resistance, and $C$ is the capacitance.
At low frequencies the impedance of the capacitor is high, and the Bode diagram demonstrates hyperbolic behavior.
At high frequencies the impedance of the capacitor is low, and the total impedance is determined by the resistance $R$ and it does not depend on frequency.
The Nyquist admittance diagram of the serial RC-circuit has a semi-circle shape (see Table~1) with a radius $1/R$ and centered at point $1/(2R)+i \cdot 0$, which easy to show by taking the magnitude of the quantity $Y-1/(2R)$.
Thus, if the Nyquist admittance plot of a measured sample has a semi-circle shape, it is highly likely that the sample can be represented as a serial RC-circuit. 
A question can arise: why for a serial RC-circuit we consider admittance locus instead impedance locus, which has a much simpler shape (the line parallel to the imaginary axis at the distance $R$)? 
The answer to this question is as follows: the admittance representation is very useful in case of the parallel connection of $N$ serial RC-circuits (see last line in Table~1). 
Indeed, the admittance locus of such system is a vector sum of all RC-circuit loci, and, as a result, the total locus will have a shape of $N$ overlapped semicircles.
The analogical shape will be observed for impedance of the serial connected parallel RC-circuits.
So, the Nyquist plot allows us to determine how many  RC-circuits we should take into account to construct the equivalent circuit. 
Thus, the immittance plotting can be a key for choosing the best equivalent circuit topology. 

When the equivalent circuit for the sample is constructed, its parameters (ratings of resistance, capacitance, \textit{etc}) should be tuned to match the equivalent-circuit spectrum to the experimentally observed spectrum. To perform this task three methods are commonly used: geometric method \cite{Tsai}, algebraic method (AM) \cite{MacdonaldD}, and complex non-linear least squares (CNLS) method \cite{Tsai,MacdonaldR,macdonald1982applicability}.
In geometric approach, which was used in the early EIS studies, the graphical operations with the immittance locus are performed to evaluate equivalent circuit parameters.
For a more accurate algebraic approach, the parameters of the equivalent circuit are calculated based on the characteristics of immittance spectrum, \textit{i.e.} the numerical parameters such as minima, maxima, inflection points, \textit{etc} are analyzed and used for the calculation of equivalent circuit parameters.
For the most widely used and simple equivalent circuits, their immittances, Nyquist plots, and algebraic method solutions are presented in Table~1.

\begin{table}[hbt!]
\caption{Widely used equivalent circuits and their immittances. Here $\omega_{\max}$ is the angular frequency at which the admittance imaginary part reaches its maximum value; $\omega_{\min}$ is the angular frequency at which the impedance imaginary part reaches its minimum value; $f_{\min}$ is the angular frequency at which its impedance magnitude reaches the minimum value; $f_{\max}$ is the frequency at which the admittance imaginary part reaches its maximum value, $\parallel$ denotes impedance of two parallel elements. 
\label{tab:Immittances}}

\centering

\begin{tabular}{cccc}
\hline
Circuit & Immittance & Nyquist plot& AM solution \\\hline
\begin{tikzpicture} [circuit ee IEC, thick]
\draw [->,color=white] (-1,0)--(1,0);
\draw [->,color=white] (0,-1)--(0,1);
\node at (0,1) {Serial RC};
\draw (-1.5,0) to [contact] (-1.5,0);
\draw (-1.5,0) to [resistor={info'={ $R$}}] (0,0);
\draw (0,0) to [capacitor={info'={ $C$}}] (0.7,0);
\draw (0.7,0) to [contact] (0.7,0);
\end{tikzpicture}&
\begin{tikzpicture} 
\draw [->,color=white] (-1,0)--(1,0);
\draw [->,color=white] (0,-1)--(0,1);
\node at (0,0) {$Z=R+\dfrac{1}{i \omega C}$};
\end{tikzpicture}
&
\begin{tikzpicture} 
\node at (0.4,1.8) {\small $Y^{\backprime\backprime}$};
\node at (1.8,-0.4) {\small $Y^{\backprime}$};
\draw [->,color=white] (0,0)--(0,2.3);
\draw [->] (-0.5,0)--(2.2,0);
\draw [->] (0,-0.5)--(0,2.2);
\draw[red, ultra thick] (0,0) arc (0:-180:-1);
\end{tikzpicture}
&
\begin{tikzpicture} 
\node at (0,0) {$\begin{aligned}
R=\dfrac{1}{2 Y^{\backprime \backprime}(\omega_{\max})},\\ \\
C=\dfrac{2 Y^{\backprime \backprime}(\omega_{\max})}{\omega_{\max}}.
\end{aligned}
$};
\end{tikzpicture}\\ \hline
\begin{tikzpicture} [circuit ee IEC, thick]
\def\x{-0.4}
\
\draw [->,color=white] (-1,0)--(1,0);
\draw [->,color=white] (0,-1)--(0,1);
\node at (0,1) {Parallel RC};
\draw (-1.5,0) to [contact] (-1.5,0);
\draw (-0.75+\x,0.5) to [resistor={info'={ $R$}}] (0.75+\x,0.5);
\draw (-0.75+\x,-0.5) to [capacitor={info'={ $C$}}] (0.75+\x,-0.5);
\draw (0.7,0) to [contact] (0.7,0);
\draw (0.75+\x,0.5)--(0.75+\x,-0.5);
\draw (-0.75+\x,0.5)--(-0.75+\x,-0.5);
\draw (-1.5,0)--(-0.75+\x,0);
\draw (0.7,0)--(0.75+\x,0);
\end{tikzpicture}&
\begin{tikzpicture}
\draw [->,color=white] (-1,0)--(1,0);
\draw [->,color=white] (0,-1)--(0,1);
\node at (0,0) {$Y=i \omega C+\dfrac{1}{R}$};
\end{tikzpicture}
&
\begin{tikzpicture} 
\node at (0.4,0.8) {\small $Z^{\backprime\backprime}$};
\node at (1.9,0.4) {\small $Z^{\backprime}$};
draw [->,color=white] (0,0)--(0,2);
\draw [->] (-0.5,0)--(2.2,0);
\draw [->] (0,-1.5)--(0,1);
\draw[red, ultra thick] (0,0) arc (0:180:-1);
\end{tikzpicture}
&\begin{tikzpicture} 
\node at (0,0) {$\begin{aligned}
R=-2 Z^{\backprime \backprime}(\omega_{\min}),\\ \\
C=-\dfrac{1}{Z^{\backprime \backprime}(\omega_{\min})\omega_{\min}}.
\end{aligned}
$};
\end{tikzpicture}\\ \hline
\begin{tikzpicture} [circuit ee IEC, thick]
\draw [->,color=white] (-1,0)--(1,0);
\draw [->,color=white] (0,-1)--(0,1);
\node at (0,1) {Serial RLC};
\draw (-1.5,0) to [contact] (-1.5,0);
\draw (-1.5,0) to [resistor={info'={ $R$}}] (0,0);
\draw (0,0) to [capacitor={info'={ $C$}}] (0.7,0);
\draw (0.7,0) to [inductor={info'={ $L$}}] (2.2,0);
\draw (2.2,0) to [contact] (2.2,0);
\end{tikzpicture}&

\begin{tikzpicture}
\draw [->,color=white] (-1,0)--(1,0);
\draw [->,color=white] (0,-1)--(0,1);
\node at (0,0) {$Z=R+\dfrac{1}{i \omega C}+i \omega L$};
\end{tikzpicture}
&
\begin{tikzpicture} 
\node at (0.4,1.5) {\small $Y^{\backprime\backprime}$};
\node at (2.2,-0.4) {\small $Y^{\backprime}$};
\draw [->] (-0.5,0)--(2.2,0);
\draw [->] (0,-1.5)--(0,1.5);
\draw[red, ultra thick] (0,0) arc (0:360:-1);
\end{tikzpicture}
&
\begin{tikzpicture} 
\node at (0,0) {$\begin{aligned}
R=|Z(f_{\min})|,\\
L=\frac{Rf_{\max}}{2\pi(f_{\min}^2-f_{\max}^2)},\\
C=\frac{1}{L(2 \pi f_{\min})^2}.
\end{aligned}
$};
\end{tikzpicture}
\\ \hline
\begin{tikzpicture} [circuit ee IEC, thick]
\draw [->,color=white] (-1,0)--(1,0);
\draw [->,color=white] (0,-1)--(0,1);
\node at (0,1) {Serial R-CPE};
\draw (-1.5,0) to [contact] (-1.5,0);
\draw (-1.5,0) to [resistor={info'={ $R$}}] (0,0);
\draw (0,0) -- (1.4,0);
\node [draw=black, fill=white] at (0.5,0) {CPE};
\draw (1.4,0) to [contact] (1.4,0);
\end{tikzpicture}&
\begin{tikzpicture} 
\draw [->,color=white] (-1,0)--(1,0);
\draw [->,color=white] (0,-1)--(0,1);
\node at (0,0) {$Z=R+\dfrac{1}{W(i \omega )^\alpha}$};
\end{tikzpicture}
&
\begin{tikzpicture} 
\node at (0.4,1.8) {\small $Y^{\backprime\backprime}$};
\node at (2.2,0.4) {\small $Y^{\backprime}$};
\draw [->] (-0.5,0)--(2.2,0);
\draw [->] (0,-0.5)--(0,2.2);
\draw[blue,dashed] (0,0) arc (-30:-390:-1);
\draw[red, ultra thick] (0,0) arc (-30:-150:-1);
\end{tikzpicture}
&\begin{tikzpicture} 
\node at (0,0) {$\begin{aligned}
R=\dfrac{1}{2 Y^\backprime (\omega_{\max})},\\
\alpha=\frac{4}{\pi}\arctan [2 R Y^{\backprime \backprime}(\omega_{\max} )],\\
W=\frac{1}{R \omega^{\alpha}_{\max}}.
\end{aligned}$};
\end{tikzpicture}\\
\hline
\begin{tikzpicture} [circuit ee IEC, thick]
\draw [->,color=white] (-1,0)--(1,0);
\draw [->,color=white] (0,-1)--(0,1);
\node at (0,1) {Two serial RC in parallel};
\draw (-1.5,0) to [resistor={info={ $R_1$}}] (0,0);
\draw (0,0) to [capacitor={info={ $C_1$}}] (0.7,0);
\draw (-1.5,-1.2) to [resistor={info'={ $R_2$}}] (0,-1.2);
\draw (0,-1.2) to [capacitor={info'={ $C_2$}}] (0.7,-1.2);
\draw (-1.5,0) -- (-1.5,-1.2);
\draw (0.7,0) -- (0.7,-1.2);
\draw (-1.5,-0.6) -- (-1.8,-0.6);
\draw (-1.8,-0.6) to [contact] (-1.8,-0.6);
\draw (0.7,-0.6) -- (1,-0.6);
\draw (1,-0.6) to [contact] (1,-0.6);
\end{tikzpicture}&
\begin{tikzpicture} 
\draw [->,color=white] (-1,0)--(1,0);
\draw [->,color=white] (0,-1)--(0,1);
\node at (0,0) {$\begin{aligned}Z_1=R_1+\dfrac{1}{i \omega C_1},\\Z_2=R_2+\dfrac{1}{i \omega C_2},\\Z=Z_1 \parallel Z_2.\end{aligned}$};
\end{tikzpicture}
&
\begin{tikzpicture} 
\node at (0.4,1.8) {\small $Y^{\backprime\backprime}$};
\node at (2.2,0.4) {\small $Y^{\backprime}$};
\draw [->] (-0.5,0)--(2.2,0);
\draw [->] (0,-0.5)--(0,2.2);
\draw[red, ultra thick] plot[smooth, color=blue] file{two_arc.dat};
\end{tikzpicture}
&\begin{tikzpicture} 
\node at (0,1) {$\begin{matrix}\mbox{Simple AM solution}\\\mbox{does not exist}\end{matrix}$};
\end{tikzpicture}\\
\hline
\end{tabular}

\end{table}

Even though the algebraic approach provides more accurate results than the geometric method, it is applicable only to simple circuits.
Generally, no exact algebraic solution exists for equivalent circuit parameters calculation (\textit{e.g.}, see last line in Table~1).
For this reason, the complex non-linear least squares method, proposed in 1978 by J.~Ross~Macdonald \cite{macdonald1978theory}, is usually applied for EIS spectrum analysis.
The key idea of the CNLS protocol is to approximate the equivalent circuit parameters by the minimization of the following functional

\begin{equation}
\label{Eq:CNLS}
    \sum\limits_{n=1}^L \left| Y_{exp}(\omega_n)-Y_{EC}(\omega_n, \vec{P})\right|^2=\min,
\end{equation}
where $Y_{exp}(\omega_n)$ is the experimentally obtained admittance at frequency $\omega_n$, $Y_{EC}(\omega_n,\vec{P})$ is the equivalent circuit admittance at frequency $\omega_n$, $\vec{P}$ is a vector of the equivalent circuit parameters (resistances, capacitances, inductions \textit{etc.}), and $L$ is a number of frequencies used to perform the measurements. The admittance in equation \eqref{Eq:CNLS} can be replaced by impedance, the magnitude of impedance, or any other immittance function. Also the weighting function $W(\omega_n)$ can be added if some frequency range has higher importance than others
\begin{equation}
    \sum\limits_{n=1}^L \left| Y_{exp}(\omega_n)-Y_{EC}(\omega_n, \vec{P})\right|^2 W(\omega_n)=\min.
\end{equation}
For the minimization of function \eqref{Eq:CNLS} a large number of optimization algorithms exist.
For example, freeware \textsc{levm} program by J.~Ross~Macdonald uses the Levenberg-Marquardt algorithm \cite{ranganathan2004levenberg}; 
\textsc{nelm} package for \textsc{matlab} and \textsc{python}, which was developed by the authors, uses the Nelder-Mead algorithm \cite{NM} and can support big data sets (the package can be obtained by a request).
It is interestingly to note, that as a starting vector $\vec{P}$, these optimization algorithms frequently use the results of the algebraic method approximation.

\subsection{Immittance models for bioelectronics}
The described above equivalent circuit concept opens one of the fruitful ways for studying biological systems.
Particularly, for contact between the metallic electrode, electrolyte (physiological medium), and biological matter (cells, tissues, macromolecules) -- the key element existing in the most number of the bio-impedance based devices -- it is possible to construct the relatively simple EC, which correctly reflect  physical, chemical, and biological phenomena in such interface. In this section we will derive this EC and demonstrate its advantages for cell research.
In order to make the explications consistent, we will firstly consider a model of a metal/electrolyte interface.
\subsubsection{Double electrical layer.}
\begin{figure}
    \centering
\subfigure[]{\begin{tikzpicture}
\draw [fill=ProcessBlue!20,draw=ProcessBlue!20] (-1,1.95) rectangle (7,-1.95);
\node at (-1,0) [rotate=90,fill=gray!20,draw=gray!50] {\huge ~Electrode~};
\node at (0,0) [circle,draw=blue!50,fill=blue!20] {\tiny +};
\node at (1,1.5) [circle,draw=blue!50,fill=blue!20] {\tiny +};
\node at (1.5,-1.5) [circle,draw=blue!50,fill=blue!20] {\tiny +};
\node at (2,0) [circle,draw=blue!50,fill=blue!20] {\tiny +};
\node at (0,1) [circle,draw=Yellow,fill=Yellow!50] {\tiny \textbf{--}};
\node at (0.5,-1) [circle,draw=Yellow,fill=Yellow!50] {\tiny \textbf{--}};
\node at (2,1.2) [circle,draw=Yellow,fill=Yellow!50] {\tiny \textbf{--}};
\node at (1.2,0.5) [circle,draw=Yellow,fill=Yellow!50] {\tiny \textbf{--}};

\node at (5.5,-0.75) [circle,draw=blue!50,fill=blue!20] {\tiny +};
\node at (6,1.5) [circle,draw=blue!50,fill=blue!20] {\tiny +};
\node at (4,-1.6) [circle,draw=blue!50,fill=blue!20] {\tiny +};
\node at (3.7,0.65) [circle,draw=blue!50,fill=blue!20] {\tiny +};
\node at (4.2,1.3) [circle,draw=Yellow,fill=Yellow!50] {\tiny \textbf{--}};
\node at (4,-0.6) [circle,draw=Yellow,fill=Yellow!50] {\tiny \textbf{--}};
\node at (6,0.7) [circle,draw=Yellow,fill=Yellow!50] {\tiny \textbf{--}};
\node at (6.5,-1.6) [circle,draw=Yellow,fill=Yellow!50] {\tiny \textbf{--}};
\draw [->] (-1.2,-2.7) -- (7,-2.7) node [very near end, below] {$x$};
\end{tikzpicture}}\\\subfigure[]{\begin{tikzpicture}
\draw [fill=ProcessBlue!20,draw=ProcessBlue!20] (-1,1.95) rectangle (7,-1.95);
\node at (-1,0) [rotate=90,fill=gray!20,draw=gray!50] {\huge ~Electrode~};
\node at (0,0) [circle,draw=blue!50,fill=blue!20] {\tiny +};
\node at (0,1.5) [circle,draw=blue!50,fill=blue!20] {\tiny +};\node at (0,-1.5) [circle,draw=blue!50,fill=blue!20] {\tiny +};

\node at (0,0.75) [circle,draw=blue!50,fill=blue!20] {\tiny +};
\node at (0,-0.75) [circle,draw=blue!50,fill=blue!20] {\tiny +};

\node at (0,0) [circle,draw=blue!50,fill=blue!20] {\tiny +};
\node at (1,-1.3) [circle,draw=blue!50,fill=blue!20] {\tiny +};
\node at (1,-0.3) [circle,draw=blue!50,fill=blue!20] {\tiny +};
\node at (1,1.2) [circle,draw=blue!50,fill=blue!20] {\tiny +};
\node at (1,0.5) [circle,draw=blue!50,fill=blue!20] {\tiny +};

\node at (2,0.6) [circle,draw=blue!50,fill=blue!20] {\tiny +};
\node at (2,-0.8) [circle,draw=blue!50,fill=blue!20] {\tiny +};
\node at (2,1.4) [circle,draw=Yellow,fill=Yellow!50] {\tiny \textbf{--}};

\node at (2.8,0) [circle,draw=blue!50,fill=blue!20] {\tiny +};
\node at (2.9,-1) [circle,draw=blue!50,fill=blue!20] {\tiny +};
\node at (3.1,0.8) [circle,draw=Yellow,fill=Yellow!50] {\tiny \textbf{--}};

\node at (5.5,-0.75) [circle,draw=blue!50,fill=blue!20] {\tiny +};
\node at (6,1.5) [circle,draw=blue!50,fill=blue!20] {\tiny +};
\node at (4,-1.6) [circle,draw=blue!50,fill=blue!20] {\tiny +};
\node at (4.2,0.65) [circle,draw=blue!50,fill=blue!20] {\tiny +};
\node at (4.2,1.4) [circle,draw=Yellow,fill=Yellow!50] {\tiny \textbf{--}};
\node at (4,-0.6) [circle,draw=Yellow,fill=Yellow!50] {\tiny \textbf{--}};
\node at (6,0.6) [circle,draw=Yellow,fill=Yellow!50] {\tiny \textbf{--}};
\node at (6.5,-1.6) [circle,draw=Yellow,fill=Yellow!50] {\tiny \textbf{--}};
\draw [->] (-1.2,-2.7) -- (7,-2.7) node [very near end, below] {$x$};
\draw[dashed] (0.5,2) --(0.5,-2);
\node at (2.5,-2.3) {$\underbrace{~~~~~~~~~~~~~~~~~~~~~~~~~~~~~~}_{\ell_D}$};
\node at (0, 2.3) {HL};
\node at (3.5, 2.3) {Gouy-Chapman layer};

\end{tikzpicture}}
    \caption{Sketch for the Gouy-Chapman-Stern problem (not in scale). (a) The ions distribution before applying a voltage to the electrode. (b) The ions distribution after applying a negative voltage to the electrode: ions form the Helmholtz layer (HL) and Gouy-Chapman layer (diffusion layer). For low electrode voltages in the Gouy-Chapman layer on the distance $\ell_D$ (Debye screening length)  the ions non-uniform concentration falls in $e=2.71828...$ times. Note, that the Helmholtz layer could be presented as several ions and solvent molecules layers. \label{fig:GC_Problem} }

\end{figure}
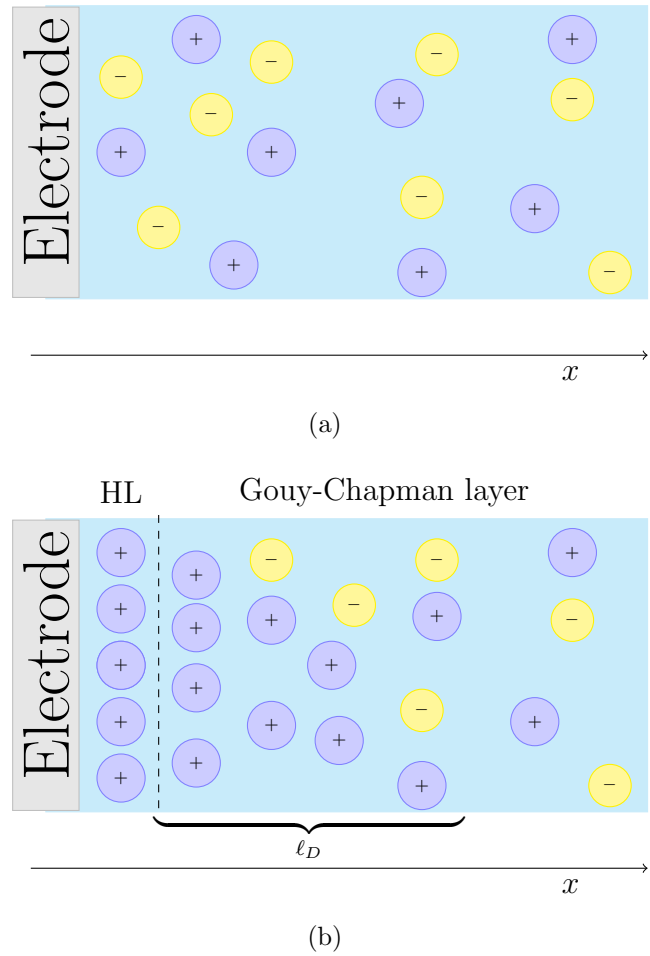

In 1853, Hermann von Helmholtz has found that in the absence of electrochemical reactions the metal/electrolyte interface, which he called  a double electrical layer (DL) \cite{Bard_book_about_DL}, demonstrates capacitor-like behavior.
Later,  Louis~G.~Gouy (1910)  and  David L. Chapman (1913)  made the first effort to construct a model of the metal/electrolyte interface, which we will briefly discuss now. For simplicity we will deal with one-dimension case.
Let's consider an planar electrode embedded in an electrolyte solution with ion concentration $\eta_0$, dielectric permeability $\varepsilon$, and temperature $T$; ions in the electrolyte are considered to be the \textit{ideal gas} [Fig. \ref{fig:GC_Problem}(a)].
What happens if the voltage $V_0$ is applied to the electrode [Fig. \ref{fig:GC_Problem}(b)]?

According to the Poisson equation, the electric potential distribution in the electrolyte can be described as 

\begin{equation}
\label{Eq:Poisson}
    \frac{\partial^2}{\partial x^2} \varphi(x)=-\frac{4 \pi}{\varepsilon}\rho(x),
\end{equation}
where $x$ is coordinate in the direction perpendicular to the electrode, and $\rho(
x)$ is a charge density formed by ions.
Besides, the distribution of ion concentration should satisfy the Boltzmann statistics, namely distribution of positively charged ions is equal to

\begin{equation}
\label{Eq:Boltzmann_1}
    \eta_{+}(x)=\eta_0 e^{-\el \varphi(x)/kT},
\end{equation}
and distribution of negatively charged ions is equal to 
\begin{equation}
\label{Eq:Boltzmann_2}
    \eta_{-}(x)= \eta_0 e^{+\el \varphi(x)/kT},
\end{equation}
where $\el$ is electron charge (here and further for simplicity we will consider singularly charged  only).
The total charge density equals 
\begin{equation}
 \label{Eq:Boltzmann}
 \rho(x)=\el \eta_{+}(x)-\el \eta_{-}(x).
\end{equation}
Combining Eqs. \eqref{Eq:Poisson}, \eqref{Eq:Boltzmann_1}, \eqref{Eq:Boltzmann_2}, and \eqref{Eq:Boltzmann} we obtain the equation for the potential

\begin{equation}
\label{Eq:Debye_Eq}
      \frac{\partial^2}{\partial x^2} \varphi(x)=\frac{8 \pi 
     \el}{\varepsilon}
     \eta_0 \sinh \frac{\el \varphi(x)}{kT}.
\end{equation}

To solve this equation, we consider the case $\el \varphi \ll kT$ (Debye approximation) and expand the right-hand side of the equation \eqref{Eq:Debye_Eq} into the Taylor series.
As a result, in the Debye approximation, the potential distribution can be written as

\begin{equation}
\label{eq:Debye_Solution}
\varphi=V_0 e^{-x/\ell_D},
\end{equation}
where 
\begin{equation}
\ell_D=\sqrt{\frac{\varepsilon kT}{8 \pi \el^2 \eta_0}}
\end{equation}
is the Debye screening length (here we assumed that $\varphi(\infty)=0$). 
For example, in the physiological solution (0.9\% NaCl) $\ell_D=8$ \AA. 
Now the electrode capacitance can be easily calculated as the charge-to-voltage ratio

\begin{equation}
\label{Eq:C_Debye}
C=\frac{Q}{V_0}=\dfrac{\int\limits_0^\infty \rho dx}{V_0}S=\dfrac{\varepsilon\varphi'(x)\Big|_0^{\infty}}{4 \pi V_0}S=\frac{\varepsilon S}{4 \pi \ell_D},
\end{equation}
where $Q$ is the total charge of the electrode, $S$ is the area of the electrode. 
It is worth noticing, that equation \eqref{Eq:C_Debye} is similar to the equation that describes the capacitance of the parallel-plate capacitor with the dielectric permeability $\varepsilon$ and inter-plate distance $\ell_D$.
Verification of the equation \eqref{Eq:C_Debye} by experimental data revealed that this model overestimates the experimental capacitance value.
To obtain more accurate results, Gouy and Chapman found the exact solution of the one-dimension equation \eqref{Eq:Debye_Eq} and used it to derive the differential capacitance of the metal/electrolyte interface 

\begin{equation}
\label{Eq:C_Gouy_Chapman}
C_{GC}=\frac{\varepsilon S}{4 \pi \ell_D}\cosh\frac{\el V_0}{2kT},
\end{equation}
which matches the Debye-approximation capacitance for $\el V_0 \ll 2kT$ [the  Eq. \eqref{Eq:Debye_Eq} could be easily integrated by multiplying it on $2\varphi'(x)$]. 
Unfortunately, already for relatively small values of voltage applied to the electrode, the Gouy-Chapman capacitance becomes too large and loses any physical meaning.

This paradox arises from the fact that the ideal gas model for ions in electrolyte is not valid  if high voltage is applied to the electrode. Indeed, from Eqs. \eqref{Eq:Boltzmann_1} and \eqref{Eq:Boltzmann_2} it is easy to see that ions density at electrode is equal
\begin{equation}
\label{eq:Density_at_surface}
    \eta(0)=\eta_+(0)+\eta_-(0)=2\eta_0\cosh\left(\frac{\el V_0}{kT}\right).
\end{equation}
On the other hand, we can estimate the maximum density of the ions as
\begin{equation}
\label{eq:Max_density}
    \eta_{\max} \approx \frac{1}{\mbox{Single ion volume}} = \frac{3}{4 \pi r^3},
\end{equation}
where $r$ is the ion effective radius. From these equations, it is clear that for NaCl 0.9\% solution at room temperature the sodium ions density at electrode $ \eta(0)$ became greater than maximum sodium ion density $\eta_{\max}$ when electrode voltage became -0.15 V (here we have used value 0.18 nm for sodium-ion radius \cite{hummer1997ion}). The similar estimation could be done if we will take as $\eta_{\max}=8/(5.65 \mbox{~\AA})^3$ -- the ``concentation'' of the ions in the crystal NaCl (the ration of atoms number per  lattice cell to lattice cell volume).  Moreover, ions radius \cite{hummer1997ion} and crystal NaCl lattice parameter (5.65 \AA \cite{Crystallography}) in this example system are comparable with Debye screening length, and thus the potential distribution \eqref{eq:Debye_Solution} and followed from it ions density became loosely applicable at $x=0$. That's why it is impossible to consider ions as point-like objects at the electrode surface and at high electrode voltages.

For this reason, ions cannot approach arbitrarily closely to the electrode, but they form a so-called \textit{Helmholtz layer} (or plane of closest approach)  on its surface.
Nevertheless, at relatively large distances from the electrode (a few $\ell_D$) ions distribution can be correctly described by the Debye approximation. 

These effects was taken into account by Otto Stern in 1924, who proposed the two-region structure for metal/electrolyte interface \cite{Bard_book_about_DL,stern1924theory,  delahay1965double}. In this model, ions are assumed to form both the \textit{Helmholtz layer} and the more distant \textit{Gouy-Chapman layer}, which is also called \textit{diffusion layer} [Fig. \ref{fig:GC_Problem}(b)]. 

The double electrical layer in Gouy-Chapman-Stern theory can be described by the equivalent circuit that contains a serial connection of the Helmholtz layer capacitance and the Gouy-Chapman capacitance. 
Due to the fact that the value of Helmholtz capacitance is usually (\textit{e.g.} at high ion concentrations) much lower than the value of the Gouy-Chapman capacitance, the capacitance of the metal/electrolyte interface is determinated by the Helmholtz layer \cite{Bard_book_about_DL,delahay1965double}.
Typically Helmholtz layer capacitance  is defined as

\begin{equation}
    C_H=S\times c_{u},
\end{equation}
where $S$ is area, and $c_u$ is experimentally obtained capacitance per area unit.
Therefore, the total immittance of the metal-electrolyte interface can be represented by a serial RC-circuit

\begin{equation}
\label{Eq:Z_Debye_Gouy_Chapman}
    Z_{m/e}=R_{b}+\frac{1}{ i \omega C_{dl}},
\end{equation}
where $R_b$ is the electrolyte bulk resistance and $C_{dl}=C_H \parallel C_{GC}$ is the double layer capacitance.
\subsubsection{Dispersion of the Double Layer Capacitance.}

Even though the Gouy-Chapman-Stern theory is completely consistent, the frequency dependence of the metal/electrolyte immittance described by Eq. \eqref{Eq:Z_Debye_Gouy_Chapman} is rarely observed due to the effect of so-called \textit{capacitance dispersion} \cite{About_C_Dispersion,About_C_Dispersion_2}.
This phenomena predict that instead capacitor-like behaviour \eqref{Eq:Z_Debye_Gouy_Chapman} immittance of the metal/electrolyte  interface is govern  by the following equation
\begin{equation}
\label{Eq:Z_Dispersion}
    Z_{m/e}=R_{b}+\frac{1}{W (i \omega)^\alpha},
\end{equation}
where $R_b$ is still the bulk resistance of electrolyte, $\alpha$ is the so-called non-ideality parameter, and $W$ is the pseudo-capacitance with dimension $\Omega^{-1}$Hz$^{-\alpha}$=$S s^{\alpha}$. 
The element with impedance $\dfrac{1}{W (i \omega)^\alpha}$ is directly refereed to capacitance dispersion and called the constant phase element (CPE), because its impedance phase $\dfrac{\pi \alpha}{2}$ is independent of frequency (for CPE properties see Table~1). 

Despite the fact that numerous experimental and theoretical papers are devoted to the subject, the true origin of the capacitance dispersion is still not elucidated \cite{martin2011influence, singh2013debye, birla2014theory,liu1985fractal,kerner2000origin}.
However, there exist two types of useful and illustrative models that partially explain this phenomenon.
The models of the first type are based on the diffusion and chemical effects, and the models of the second type explain capacitance dispersion via the spatial nonuniformity current flow through an electrode.
Here, we will outline the two well-known examples of such models, one for each type.

The first example is the model proposed by E. Warburg (the so called \textit{Warburg impedance}) \cite{Barsoukov}. The model can be introduced as follows.
Let's assume that after applying voltage to the electrode, the non-equilibrium ions concentration is generated on its surface.
For example, this non-equilibrium concentration can arise from the electrochemical reactions between electrolyte ions and atoms of the electrode.
Due to the fact that the ions form a non-uniform concentration $\eta(x,t)$, a diffusion flow of ions (\textit{i.e.} diffusion current) also takes place (in this model, if the Debye length is small and the electrical field in the electrolyte is weak  we can neglect the drift electrical current).

Then the assumption that the concentration of ions on the electrode surface (at $x=0$) is proportional to the magnitude of the applied voltage $V_0$, \textit{i.e.} $\eta\Big|_{x=0}=\gamma V_0$ is made. 
Thus, in the case of one-dimensional distribution of ions and the harmonic excitation voltage $V_0:=V_0 e^{i \omega t}$ we can write the following set of equations which are based on diffusion equation

\begin{equation}
    \left\{\begin{aligned}
       & D\frac{\partial^2}{\partial x^2}\eta(x, t) =\frac{\partial}{\partial t} \eta(x, t),\\
       & \eta(x,t)\Big|_{x = 0} = \gamma V_0 e^{i\omega t}, \\
       &\eta(x,t)\Big|_{x = +\infty} = 0,
    \end{aligned}\right. 
\end{equation}
where $D$ is the diffusion coefficient. 
After solving this system, we obtain the expression for the ion concentration

\begin{equation}
    \eta(x, t) = \gamma V_0 e^{-\sqrt{i \omega/ D} x} e^{i\omega t}.
\end{equation} 

Finally, to derive Warburg immittance we should use Fick's law to calculate the current through the electrode (at $x=0$) and relate it to the excitation voltage $V_0$.
As a result, we obtain

\begin{equation}
    Y_W = \frac{-
    \el D \eta'(x=0) e^{i\omega t}}{V_0e^{i\omega t}} = \el \gamma\sqrt{i\omega D}
\end{equation}
or 
\begin{equation}
\label{Eq:Warburg_Impedance}
    Z_W = \frac{1}{\el \gamma\sqrt{ D} (i\omega)^\frac{1}{2}}.
\end{equation}

From Eq. \eqref{Eq:Warburg_Impedance} one can see that taking into account the diffusion effects results in capacitance dispersion with the non-ideality parameter $\alpha=\dfrac{1}{2}$.

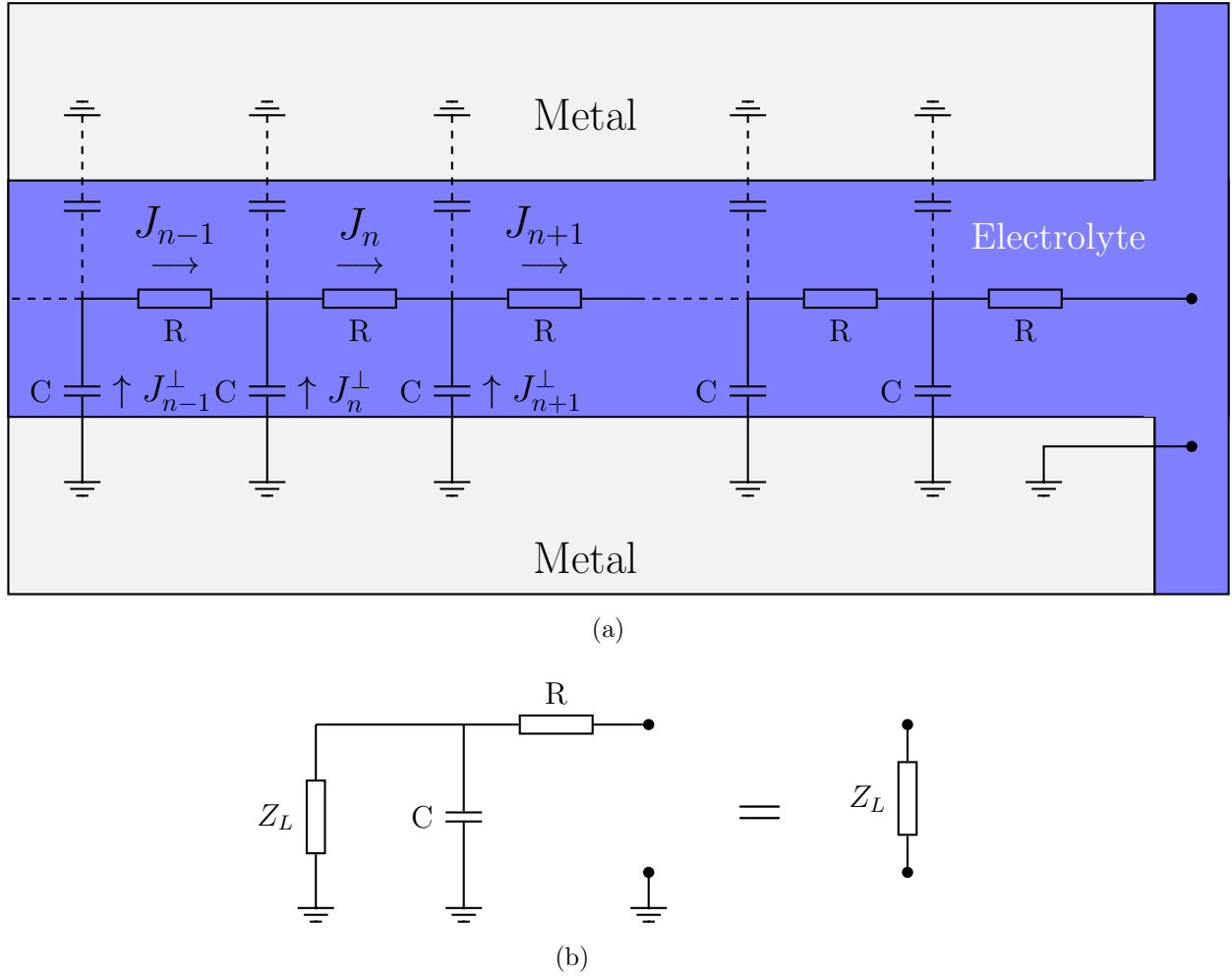
\begin{figure}
   \centering
   \subfigure[]{    
       \begin{tikzpicture}[circuit ee IEC, thick]
     \draw [fill=gray!10](-1,6.5) rectangle (14.5,-1.5);
    \draw [fill=blue!50](-1,0.9) rectangle (15.5,4.1);
     \draw [fill=blue!50](14.5,6.5) rectangle (15.5,-1.5); 
     \draw [draw=blue!50,line width=3mm] (14.5,0.9) -- (14.5,4.1);
    \draw (0,0) to [capacitor={info={C},info'={\large $\uparrow J_{n-1}^\perp$}}] (0,2.5);
    \draw (0,0) to [ground] (0,-0.15);    
    \draw (0,2.5) to [resistor={info'={R},info={\Large $\underset{\longrightarrow}{J_{n-1}}$}}] (2.5,2.5);
    \draw [dashed] (0,2.5) to [capacitor] (0,5);
    \draw [dashed] (0,5) to [ground] (0,5.15);
    \draw [dashed] (0,2.5) -- (-1, 2.5);
    \draw (2.5,0) to [capacitor={info={C},info'={\large $\uparrow J_n^\perp$}}] (2.5,2.5);
    \draw (2.5,0) to [ground] (2.5,-0.15);
    \draw (2.5,2.5) to [resistor={info'={R},info={\Large $\underset{\longrightarrow}{J_{n}}$}}] (5,2.5);
    \draw [dashed] (2.5,2.5) to [capacitor] (2.5,5);
    \draw [dashed] (2.5,5) to [ground] (2.5,5.15);
    
    \draw (5,0) to [capacitor={info={C},info'={\large $\uparrow J_{n+1}^\perp$}}] (5,2.5);
    \draw (5,0) to [ground] (5,-0.15);
    \draw (5,2.5) to [resistor={info'={R},info={\Large $\underset{\longrightarrow}{J_{n+1}}$}}] (7.5,2.5);
    \draw [dashed] (5,2.5) to [capacitor] (5,5);
    \draw [dashed] (5,5) to [ground] (5,5.15);
    
    \draw (9,0) to [capacitor={info={C}}] (9,2.5);
    \draw (9,0) to [ground] (9,-0.15);
    \draw (9,2.5) to [resistor={info'={R}}] (11.5,2.5);
    \draw [dashed] (9,2.5) to [capacitor] (9,5);
    \draw [dashed] (9,5) to [ground] (9,5.15);
    
    \draw (11.5,0) to [capacitor={info={C}}] (11.5,2.5);
    \draw (11.5,0) to [ground] (11.5,-0.15);
    \draw (11.5,2.5) to [resistor={info'={R}}] (14,2.5) -- (15,2.5) to[contact] (15,2.5);
    \draw [dashed] (11.5,2.5) to [capacitor] (11.5,5);
    \draw [dashed] (11.5,5) to [ground] (11.5,5.15);
    \draw  (15,0.5) to[contact] (15,0.5) -- (13,0.5) --  (13,0) to [ground] (13,-0.15);
    \draw[dashed] (7.5,2.5)--(9,2.5);
    
    \node at (6.8,-1) {\Large Metal};
    \node [color=white] at (13.2,3.3) {\large Electrolyte};
    \node at (6.8,5) {\Large Metal};
    \end{tikzpicture}} \\
    \subfigure[]{    \begin{tikzpicture}[circuit ee IEC, thick]
     \draw (-2,2.5) to [resistor={info'={$Z_L$}}] (-2,0);
     \draw (-2,0) to [ground] (-2,-0.15); 
     
    \draw (6,2.5) to [contact] (6,2.5) to [resistor={info'={$Z_L$}}] (6,0.5);
     \draw (6,0.5) to [contact] (6,0.5);
     
     \draw (-2,2.5) -- (0,2.5);
    \draw (0,0) to [capacitor={info={C}}] (0,2.5);
    \draw (0,0) to [ground] (0,-0.15);    
    \draw (0,2.5) to [resistor={info={R}}] (2.5,2.5) to[contact] (2.5,2.5);
    \draw (2.5,0.5) to[contact] (2.5,0.5) -- (2.5, 0) to[ground] (2.5,-0.15);
  
    \node at (4,1.25) {\huge =};
   
    \end{tikzpicture}}
    \caption{On the Levie problem. (a) An infinite pore in the electrode can be represented as an infinite ladder RC circuit. Here, voltage is applied to the metal and to the right part of the electrolyte.  The direction from the surface of the electrode into the depth of the pore is from the right to the left. (b) The impedance of such pore $Z_L$ can be easily  calculated, if we noted, that it does not change after connecting one more RC-branch.\label{Fig:Levie}}
  
    \label{fig:transmission_line}
\end{figure}

The second example that we will consider is the R. Levie model of an electrode with a scratch on its surface \cite{de1965influence}.
For simplicity, we will deal with the limiting case of the Levie model, when a scratch has infinite depth and degenerates into the pore [see Fig. \ref{Fig:Levie}(a), the direction from right to left corresponds to going deeper to pore along its central axis]. 
To calculate the immittance of this system, the so-called transmission-line approximation approach can be used. 
Let's divide the pore across its symmetry axis into parts with infinitely small length (Fig. \ref{fig:transmission_line}).
One can see that the current $J_n$ that flows through the cross-section resistance of the $n$-th section is formed by the sum of the current $J_{n-1}$ in the $(n-1)$-th section and the current  $J^\perp_n$ that flows through the metal/electrolyte impedance (which we assume to be capacitive).
The current that flows through the $n$-th section is added to the similar current $J_{n+1}$ that flows through the $(n+1)$-th section resistance, and so on ($J_{n+1}=J_n+J^\perp_{n+1},\,J_{n+2}=J_{n+1}+J^\perp_{n+2},$ \textit{etc}). 
Therefore in transmission-line approximation a pore can be modeled as an infinite ladder RC-circuit, the impedance $Z_L$ of which can be easily calculated \cite{Feynman_Lectures}.
Indeed, notice that the value of $Z_L$ does not change if we connect one more RC-branch to the circuit [Fig.~\ref{Fig:Levie}(b)]

\begin{equation}
    Z_L=R+Z_L \parallel \frac{1}{i \omega C}, 
\end{equation}
where the designation $ \parallel$ symbolizes the impedance of the two parallel elements, and thus 
\begin{equation}
\label{Eq:Levie_Z}
    Z_L=\frac{R}{2}\left(1+\sqrt{1+\frac{4}{i \omega RC}}\right).
\end{equation}
It is easy to see that in the ultra low frequency limit Eq. \eqref{Eq:Levie_Z} can be written as 

\begin{equation}
\label{Eq:Levie_Z_lim}
    Z_L=\sqrt{\frac{R}{i \omega C}},
\end{equation}
and again we observe the capacitance dispersion with $\alpha=\dfrac{1}{2}$.

\subsubsection{Metal/electrolyte/cell interface.} 
\label{Sec:Giaever-Keese_model}
Up to date, there exist a large number of models that describe the electrical properties of the metal/electrolyte/cell interface: an electrode carrying either the cell population or a single cell on its surface \cite{siddiquei2010electrical, richardson2007polymerization, rahman2008detailed, mondal2013extended, wang2008analysis}. 
One of the most popular qualitative equivalent circuits for the metal/electrolyte/cell interface is the ECIS model proposed by I. Giaever and C. Keese \cite{Geaever}.  To better understand its origin, the Table 2 lists the most significant properties of the MEC contact. 
From data given in this table, one can see that a cell itself is a very strong dielectric: at low frequencies a cell impedance is higher than the impedance of the pure metal-electrolyte interface. This picture is due to the lipid-based composition of a membrane and the very selective ion channels in the form of proteins \cite{Plonsey}.
So, in fact one can neglect current flow through cell at low frequencies.
At the same time, we can notice that cells are attached to the electrode surface by so-called focal contacts, \textit{i.e.} the most part of cell membrane does not have a direct contact with the electrode surface \cite{wierzbicki2013mapping}.
Thus, between cell's membrane and electrode surface exist an effective seal.
The Giaever-Keese model is based on this statement \cite{Geaever,cheran2008probing}. 
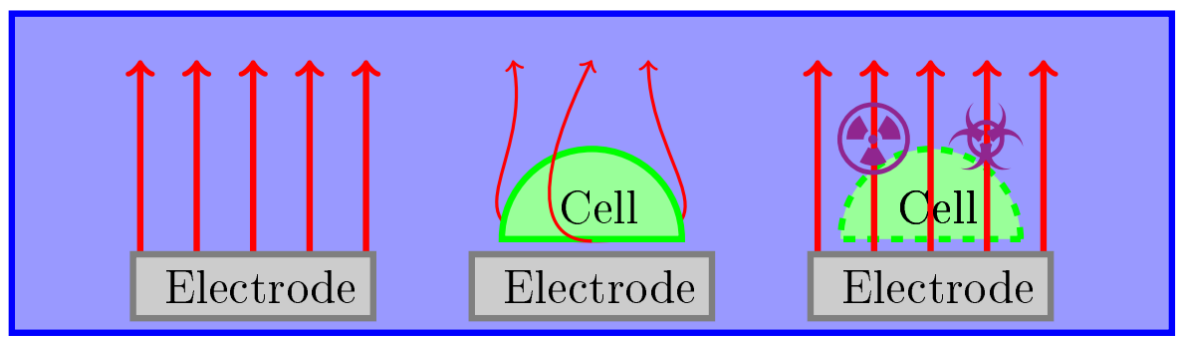
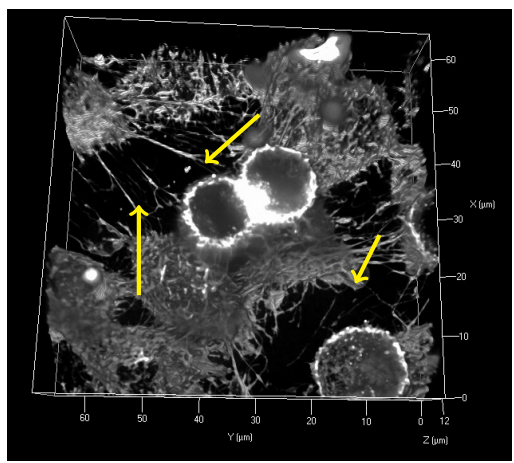
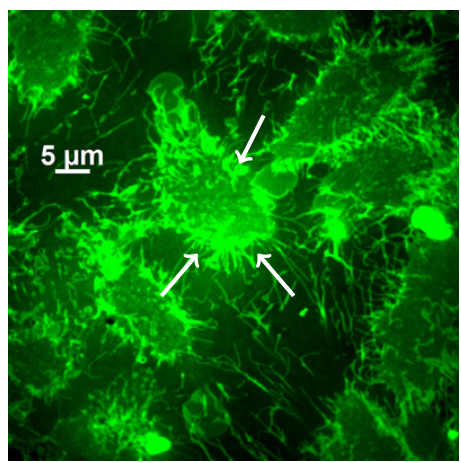
\begin{figure}
    \centering
  \subfigure[]{  \includegraphics[scale=0.5]{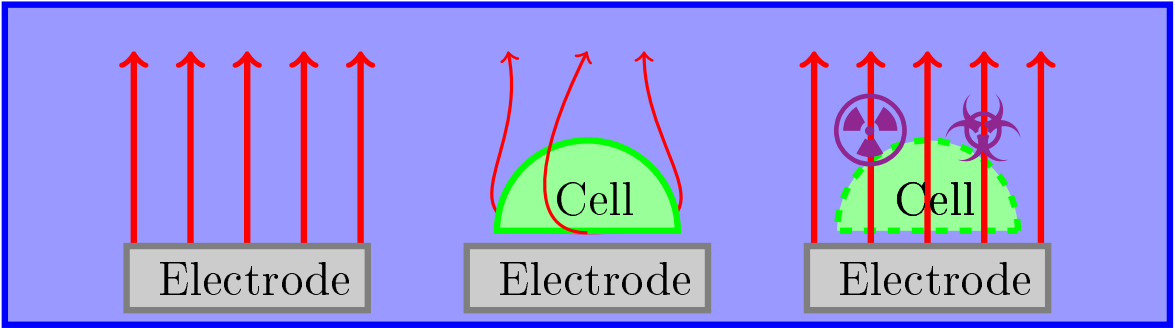}}\\

    \subfigure[]{\begin{tikzpicture}[scale=0.8]  
\node at (0,0) {\includegraphics[scale=0.4]{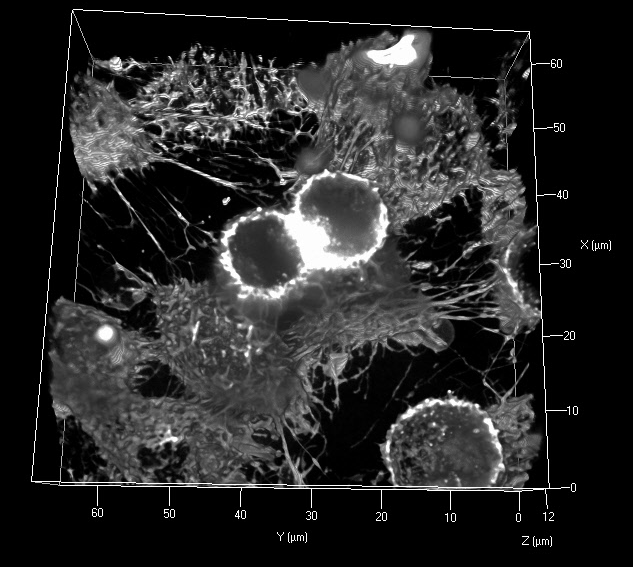}}; 
\draw[->, yellow, ultra thick] (0,2) -- (-0.9,1.2);
\draw[->, yellow, ultra thick] (2,0) -- (1.6,-0.8);
\draw[->, yellow, ultra thick] (-2,-1) -- (-2,0.5);
\end{tikzpicture}}    \subfigure[]{
\begin{tikzpicture}[scale=0.8]    
\node at (0,0) {\includegraphics[scale=0.424]{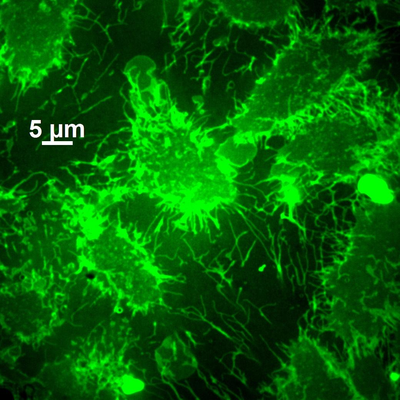}}; 
\draw[->, white, ultra thick] (1,-1) -- (0.4,-0.3);
\draw[->, white, ultra thick] (0.5,2) -- (0.1,1.2);
\draw[->, white, ultra thick] (-1.2,-1) -- (-0.6,-0.3);
\end{tikzpicture}
}
    \caption{On the Giaever-Keese model in the low-frequency limit. (a) The scheme of the metal/electrolyte/cell interface. Here, red arrows denote the electrical current. The cell-covered electrode (center panel) has a higher impedance than the empty electrode (left panel) because in the case of metal/electrolyte/cell interface current flows not only through an electrochemical impedance but also through the \textit{seal} between the cell membrane and the electrode. If a cell membrane becomes distorted upon interaction with toxins or radiation (right panel), the current has the possibility to bypass the seal through the distorted membrane, and therefore impedance spectra of cell-covered and empty electrodes become close to each other. (b, c) Morphology of HeLa cells adhesion [confocal microscopy obtained with microscope Zeiss Observer.Z1 (Zeiss, Germany) and Deep Red Cell Mask dye, pseudocolor]. Cells are attached to the surface by a small number of focal adhesion contacts (several contacts are highlighted with yellow and white arrows), and a large part of the cell membrane does not have direct contact with the surface. This phenomenon forms the above-mentioned seal. \label{Fig:Giaever_Keese}}
  
\end{figure}

In order to pass the metal/electrolyte/cell interface the current has to overcome the electrochemical impedance $Z_{\mbox{\tiny m/e}}$ [Eq. \eqref{Eq:Z_Dispersion}, Fig. \ref{Fig:Giaever_Keese}(a), left panel] and also the resistance of the seal  between the cell and the electrode surface [Fig. \ref{Fig:Giaever_Keese}(a), center panel].
This phenomenon can be described with a serial active resistance $R_{\mbox{\tiny S}}$ in the metal/electrolyte/cell interface

\begin{equation}
\label{Eq:GK-model}
    Z_{MEC}=Z_{m/e}+R_s.
\end{equation}
It should be notices, that the additional resistance $R_s$ has a geometrical nature because the current-carrying ion fluxes at low frequencies can not penetrate the cell and should flow along the cell membrane following its shape.
As far as $Z_{MEC}$ depends on the electrical and morphological properties of a cell, its value can be used for the functional characterization of cells.
For example, if cells become damaged by toxins or radiation, the distortion of the cell membrane takes place.
This results in a decrease of $R_{\mbox{\tiny S}}$ and an increase of current through metal/electrolyte/cell interface [Fig. \ref{Fig:Giaever_Keese}(a), right  panel]. 
In this case, the alteration of $Z_{\mbox{\tiny MEC}}\simeq Z_{\mbox{\tiny m/e}}$ is observed, which could used for non-optical and label-free cells viability estimation  \cite{stupin2018single,ke2011xcelligence}. 
\begin{table}[hbt!]
\centering
\begin{tabular}{lc}
\hline Quantity & Value \\ 
\hline Dielectric permeability of the membrane& 3 \cite{Plonsey}\\ 
 Cell membrane thickness & 75 \AA \cite{Plonsey}\\
Cell membrane capacitance per unit area   &$ \sim $ 100 nF/cm$^2$ \cite{Plonsey}, 0.95 $\mu$F/cm$^2$ \cite{hibino1993time}\\
Whole cell membrane capacitance for 30-$\mu$m cell   &$ \sim $ 1 pF\\
 Electrical conductivity of the cytoplasm & 220 $\Omega\, \cdot$ cm \cite{hibino1993time} \\

 Ion channel resistance & 10 G$\Omega$ \cite{Plonsey} \\
Physiological medium resistance & 50 $\Omega$ $\cdot$ cm \cite{BioImp}\\
 Typical bulk electrolyte resistance for 30-$\mu$m electrode   & 50 k$\Omega$ \cite{stupin2018single}\\
DL capacitance   (300 K, 0.9\% NaCl, 10 mV)
& 0.8 pF /$\mu$m$^2$, Eq. \eqref{Eq:C_Gouy_Chapman}\\
\hline
\end{tabular}
\caption{Electrical properties of cells and bioelectrodes.\label{tab:Cell_electrical_properties}}
\end{table} 
Also in the framework of the Giaever-Keese model, the EIS of the metal/electrolyte/cell interface can be used  to investigate cell motility \cite{szulcek2014electric}, to study interactions between cells \cite{Giaever_book}, in pharmacology and   in wound healing research \cite{asphahani2007cellular,keese2004electrical}, in biosensorics \cite{tlili2003fibroblast}, and even as a diagnostic tool in vision-prosthesis devices \cite{luo2013mri}.

\section{Experimental techniques in impedance spectroscopy}
\begin{figure}
    \centering
    \subfigure[]{\begin{tikzpicture}[thick, scale=1]
\draw[->] (0,-3)--(0,3) node [above, midway, align=center, sloped]{};
\draw[->] (0,0)--(7,0)  node [below  right, very near end]{\large $t$};
\pgfsetlinewidth {0.3 mm}
\draw[color=blue!50] plot[smooth, color=blue] file{V.dat};
\draw[color=red!50] plot[smooth, color=blue] file{J.dat};
\node at (6.5,2.9) [rectangle,fill=white, draw=black!50] {\textbf{\large \textcolor{blue!50}{$V$}~\textcolor{red!50}{$J$}}};
\end{tikzpicture}}~~~~~~\subfigure[]{\begin{tikzpicture}[thick, scale=1]
\draw[->] (0,-3)--(0,3) node [above, midway, align=center, sloped]{};
\draw[->] (0,0)--(7,0)  node [below  right, very near end]{\large $t$};
\pgfsetlinewidth {0.3 mm}
\draw[color=blue!50] plot[smooth, color=blue] file{V_r.dat};
\draw[color=red!50] plot[smooth, color=blue] file{J_r.dat};
\node at (6.5,2.9) [rectangle,fill=white, draw=black!50] {\textbf{\large \textcolor{blue!50}{$V$}~\textcolor{red!50}{$J$}}};
\end{tikzpicture}}
    \caption{The comparison of excitation voltage and current waveforms between the frequency-domain approach (a) and the time-domain approach (b). \label{Fig:Comparion_of_waveforms}}
  
\end{figure}
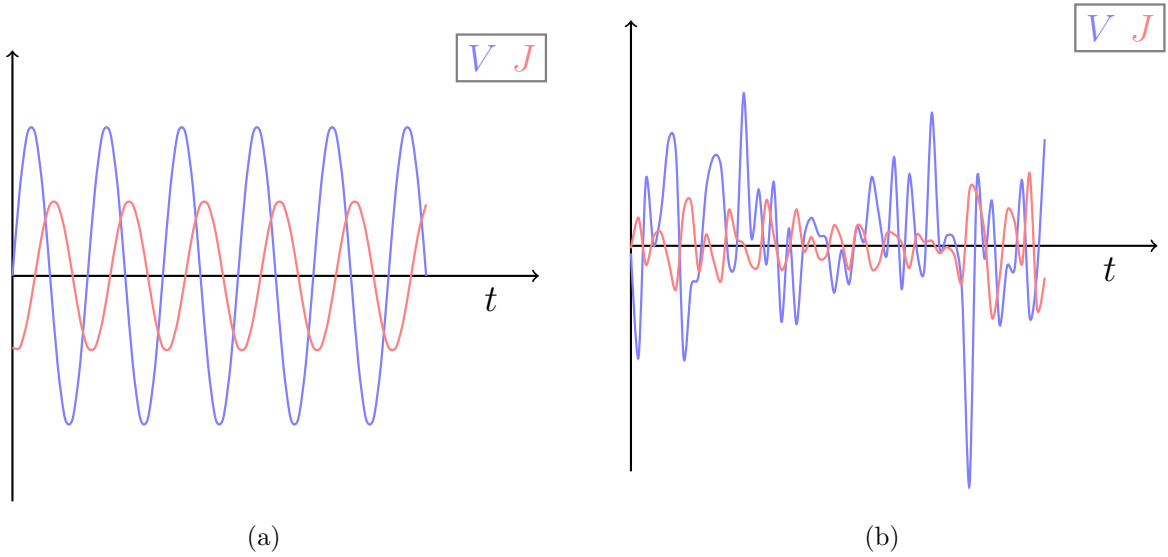

The wide range of EIS applications resulted in the appearance of a large number of methods for immittance measurements, which can be divided into frequency-domain approaches and time-domain approaches \cite{Review}. To illustrate this classification let’s consider the most common scheme for immittance measurements, which is depicted on  Fig. \ref{Fig:IS_measurement}.
Frequency-domain methods use a sine-shaped excitation voltage at a single frequency [Fig. \ref{Fig:Comparion_of_waveforms}(a)], and immittance is measured in a step-wise frequency-by-frequency manner  \cite{Barsoukov}. 
The advantages of frequency-domain approaches include high noise-immunity and the simplicity of experimental setups. 
However, a fundamental drawback of these approaches is a very low measurement rate.

In time-domain methods, the sample is scanned simultaneously at all frequencies from a broad frequency region of interest by using EV with wide spectrum \cite{Review,DFT,DFT2}.
A typical waveform for time-domain methods is presented in Fig. \ref{Fig:Comparion_of_waveforms}(b). 
Such strategy dramatically reduces the time required for a single immittance measurement. 
However, classical time-domain methods that are based on the Fourier transformation demonstrate very low noise immunity.
The combination of high measurement rate and high noise immunity can be achieved by using adaptive-filtering EIS \cite{Stupin2017}.

Below we will consider the most commonly used EIS experimental implementations \cite{Barsoukov, Lvovich,Review}. Note, that for all discussed immittance measuring approaches the EV could be recorded directly by analogue-to-digital converter (ADC), and the current response could be recorded by using ADC and ammeter, which scheme depicted in Fig. 12.5 in Ref. \cite{Tietze}.

\subsection{Frequency-domain approaches}
\subsubsection{Oscilloscopic approach.} 
The most simple way to determine immittance in the frequency-domain approach is to use a two-channel oscilloscope
with one channel connected to a voltmeter and another channel connected to an ammeter output (see Fig.~\ref{Fig:IS_measurement}) \cite{Barsoukov}. 
The data that is typically observed in the oscilloscopic approach is presented in Fig. \ref{Fig:Comparion_of_waveforms}(a).
Accordingly, the magnitude of the impedance can be derived from the ratio of voltage amplitude to current amplitude, and the phase of the impedance can be derived from the phase shift between voltage and current [see Eq. \eqref{Eq:immittance}].
The advantages of the oscilloscopic approach are the simplicity and the geometric visualization, the disadvantages include a very low accuracy and instability to noises. Oscilloscopic approach could be used in 	preliminary bio-EIS studies.

\subsubsection{Measurements with a bridge method.} 
Up to date, one of the most accurate methods for EIS measurements is based on the usage of a measurement bridge. 
\begin{figure}[h!]
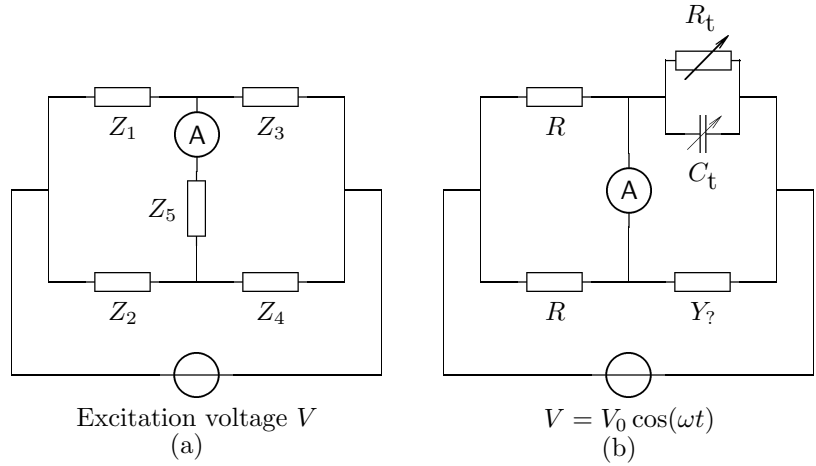

\begin{center}
\subfigure[]{
\begin{circuit}{0}
\- 5 u
\- 2 r
\nl\R1 {$ Z_1 $} r
\- 2 r
\- 2 r
\nl\R2 {$ Z_3 $} r
\- 2 r
\- 5 d
\- 5 d
\- 2 l
\nl\R2 {$ Z_4 $} l
\- 2 l
\- 2 l
\nl\R2 {$ Z_2 $} l
\- 2 l
\- 5 u
\- 2 l
\- 10 d
\- 8 r
\nl\varU1 {Excitation voltage $V$} r
\- 8 r
\- 10 u
\- 2 l
\shift -8 5
\nl\A1 {} d
\nl\R2 {$ Z_5 $} d
\- 2 d
\end{circuit}}~~~~~\subfigure[]{
\begin{circuit}{0}
\- 5 u
\- 2 r
\nl\R1 {$ R $} r
\- 2 r
\- 2 r
\- 2 d
\- 1 r
\nl\Cvar1 {$ C_{\mbox{t}} $} r
\- 1 r
\- 2 u
\- 2 r
\- 5 d
\- 5 d
\- 2 l
\nl\R2 {$ Y_? $} l
\- 2 l
\- 2 l
\nl\R2 {$ R $} l
\- 2 l
\- 5 u
\- 2 l
\- 10 d
\- 8 r
\nl\varU1 {$V=V_0\cos(\omega t)$} r
\- 8 r
\- 10 u
\- 2 l

\shift -8 5
\- 3 d
\nl\A1 {} d
\- 3 d
\shift 6 10
\- 2 u 
\nl\Rvar1 {$R_t$} . D
\shift -1 1
\- 2 d
\shift 1 4
\put {$ R_{\mbox{t}} $}

\end{circuit}}
\caption{The scheme of the Wheatstone bridge. (a) Common scheme; (b) practical realisation of the Wheatstone measuring bridge. \label{fig:Wheatstone}}
\end{center}
\end{figure}
In the simplest case, the measurement bridge is represented by the Wheatstone bridge shown in Fig. \ref{fig:Wheatstone}(a). 
Its working principle is based on an obvious fact -- the ammeter~A will show a zero-current value (no current flow through $Z_5$) when the voltage difference at $Z_5$ terminals will be equal to zero. 
This situation takes place when

\begin{equation}
\label{Eq:bridge}
    \frac{Z_1}{Z_3}=\frac{Z_2}{Z_4},
\end{equation}
which can be derived easily by using the divider relation. 

To measure the immittance with Wheatstone bridge, particularly, one can replace $Z_4$ element with investigated sample with unknown admittance $Y_?$, replace $Z_1$ and $Z_2$ elements with frequency-independent resistors (\textit{e.g.} $Z_1=Z_2=R$), replace $Z_3$ with an element that has tunable real and imaginary parts of immittance (\textit{e.g.} a parallel RC-circuit with tunable capacitor $C_{\mbox{t}}$ and tunable resistor $R_{\mbox{t}}$), short-circuit element $Z_5$, and apply to bridge single-sine EV at frequency $\omega$, for example $V=V_0\cos(\omega t)$ [Fig. \ref{fig:Wheatstone}(b)].
At the next step, one has to tune RC-circuit immittance in such a way to obtain zero current through the ammeter. 
Then the immittance of investigated sample at frequency $\omega$ can be easily calculated with Eq. \eqref{Eq:bridge}, namely
\begin{equation}
    Y_?(\omega)=\frac{1}{R_{\mbox{t}}}+i \omega C_{\mbox{t}}.
\end{equation}
It should alse be noted, that the bridge methods are stable to drift of the EV amplitude due to their zero-current measuring principle. The bridge methods due to their circuit simplicity could be useful in portable bio-EIS devices.
\subsubsection{Frequency response analyser.}
Finally, we will consider the frequency response analyser, which is currently the most popular frequency-domain EIS technique \cite{Barsoukov,Lvovich}.
The main idea of this approach is to use a lock-in amplifier \cite{Tietze}, which is extremely useful for immittance measurements. 
Indeed, let's consider the excitation voltage with the shape

\begin{equation}
\label{Eq:EV_single_sine}
V=V_0\cos(\omega t)=\Re V_0 e^{i \omega t},
\end{equation}
where $\omega$ is the angular frequency of the excitation voltage and $V_0$ is the amplitude of the excitation voltage. 
Due to Ohm's law, the current response to the excitation voltage \eqref{Eq:EV_single_sine} can be written as follows

\begin{equation}
\label{Lock-in:Current}
J=V_0|Y(\omega)|\cos\left[\omega t+\varphi(\omega)\right]=\Re\left[V_0 e^{i\omega t}|Y(\omega)| e^{i\varphi(\omega)}\right],
\end{equation}
where $|Y(\omega)|$ is the admittance magnitude and $\varphi(\omega)$ is the admittance phase. 
Now let's take into account additive current noise $\varepsilon(t)$, and multiply \eqref{Lock-in:Current} by $(2/V_0)\cos(\omega t)$, after which let's  take the average of  the result over the period of the excitation voltage $T=2 \pi / \omega$
\begin{equation}
\label{Eq:Real_Part_Lock_In}
\frac{2}{T}\int\limits_{-T/2}^{T/2} \frac{J+\varepsilon(t)}{V_0}\cos(\omega t) dt=\frac{2}{T}\int\limits_{-T/2}^{T/2} |Y(\omega)|\cos\left[\omega t+\varphi(\omega)\right]\cos(\omega t)dt=\Re Y(\omega),
\end{equation}
\textit{i.e}., we obtain the real part of the admittance. If we will make the similar procedure but at the beginning we multiply Eq. \eqref{Lock-in:Current} by  $-(2/V_0)\sin(\omega t)$ we will obtain imaginary part of the sample's admittance
\begin{equation}
\label{Eq:Imag_Part_Lock_In}
-\frac{2}{T}\int\limits_{-T/2}^{T/2}\frac{J+\varepsilon(t)}{V_0}\sin(\omega t)dt=-\frac{2}{T}\int\limits_{-T/2}^{T/2} |Y(\omega)|\cos\left[\omega t+\varphi(\omega)\right]\sin(\omega t)dt=\Im Y(\omega).
\end{equation}
Here we used a plausible assumption that noise does not correlate with the excitation voltage

\begin{equation}
\frac{2}{T}\int\limits_{-T/2}^{T/2}\varepsilon(t)\cos(\omega t)dt=0,\,\, \frac{2}{T}\int\limits_{-T/2}^{T/2}\varepsilon(t)\sin(\omega t)dt=0.
\end{equation} 

Thus, by applying the multiply-and-average technique it is possible to calculate the real and imaginary parts of the admittance independently, even in the presence of noise.
In practice,  this algorithm could be implemented in common scheme Fig. \ref{Fig:IS_measurement} by additional usage of  analogue four-quadrant multipliers  for providing product operations in Eqs. \eqref{Eq:Real_Part_Lock_In} and \eqref{Eq:Imag_Part_Lock_In} and by using  low-pass filter for providing averaging operation 
(see Sec. 11.8 and Sec. 13 in Ref. \cite{Tietze}). The lock-in technique is currently used in ECIS device \cite{Giaever2} and in the other home-made setups for cells research.

It is worth noticing that in frequency response analyzer approach it is possible to use Kramers-Kronig relations to derive the imaginary part from the real part and vice versa, \textit{i.e.} it is not necessary to measure both parts of admittance experimentally.
However, researchers usually measure both parts of the admittance and then check them for validity  by  Kramers-Kronig relations~\cite{Barsoukov,hu1989kramers}.

\subsection{Time-domain methods}
\subsubsection{Fourier EIS.}

Above we have mentioned that the key disadvantage of frequency-domain methods is a low measurement rate.
For example, it will take at least 10 seconds to measure impedance in the frequency range from 1 Hz to 40 kHz with a 1~Hz resolution. 
This problem of low measurement rate was solved in 1992 by G.~Popkirov, who  proposed  to scan the sample simultaneously in a broad frequency range by applying the broad-spectrum excitation voltage  and then to use Fourier transformation for extracting sample's immittance  from measured EV and current response time-domain sequences \cite{DFT} [a typical waveform of EV for this approach is depicted on Fig. \ref{Fig:Comparion_of_waveforms}(b)]. \textit{I.e.} sample's immittance could be derived as a relation between the Fourier images of excitation voltage and  current responses. Such method is called Fourier-EIS \cite{Review, DFT, DFT2}.

The idea of Fourier-EIS is inspired by the Eq. \eqref{Eq:Current}. 
In practice however, a discrete Fourier transformation \cite{Brigham} is used rather than a continuous transformation, because excitation voltage and current response are recorded with an analogue-to-digital converter and stored in computer memory as discrete sequences  $V_k$ and $J_k$ of the excitation voltage and current response values respectively ($k$ is the time counter).
Thus, because any periodical sequences can be expanded into discrete Fourier series, immittance on the frequency $f_m$ can be calculated as

\begin{equation}
\label{Ohm_Fourier}
Z_m=\frac{\tilde{J}_m}{\tilde{V}_m},\, Y_m=\frac{\tilde{V}_m}{\tilde{J}_m},
\end{equation}
where

\begin{equation}
\label{V_DFT}
\tilde{V}_m=\frac{1}{N}\sum\limits^{N-1}_{j=0}V_k e^{2 \pi i f_m \cdot k (T/N)},
\end{equation}
\begin{equation}
\label{J_DFT}
\tilde{J}_m=\frac{1}{N}\sum\limits^{N-1}_{j=0}J_k e^{2 \pi i f_m \cdot k(T/N)}
\end{equation} 
are discrete Fourier images of the excitation voltage and current response respectively, $N$ is the length of the recorded excitation voltage and current response sequences, $T$ is the duration of the measurement, $f_m=m/T$ is the discrete Fourier harmonic frequency ($m\in \mathbb{Z}$, $0 \leq m \leq N$). 
Thereby, the resolution of the Fourier-EIS method is $\dfrac{1}{T}$. 

It is obvious that the minimal measurement rate required for Fourier-EIS is limited by the period of the lowest-frequency harmonic of the excitation voltage.
So, for mentioned above example with Fourier-EIS it will take 1 second to measure impedance in the frequency range from 1 Hz to 40 kHz with a 1 Hz resolution, \textit{i.e.} time-domain Fourier EIS could be  faster on the order with respect to frequency-domain methods. 
Finnally it should be noticed, that Fourier-EIS is most efficient in case when $N=2^d$, $d \in \mathbb{N}$, because in this case it is possible to use fast Fourier transformation algorithms that significantly accelerate Fourier image \eqref{V_DFT} and \eqref{J_DFT} calculations \cite{Brigham}. 
Today Fourier-EIS is used for investigating fast processes in the cells \cite{matsumura2018dependence}.
\subsubsection{Adaptive filtering based EIS.}

Unfortunately, a high measurement rate of Fourier-EIS is in direct contradiction with a classical noise cancellation technique that performs signal accumulation and averaging (see, for example \cite{Statistics_book}).
Thus, if an apparatus noise cancellation method such as shielding, temperature decreasing, varying electrode geometry, \textit{etc.} are insufficient (which frequently takes place in \textit{bio-applications}), then the current response and thus the Fourier-EIS immittance spectrum will be dramatically corrupted with noises.

In rare cases such problem can be partially solved with the CNLS method. 
However, in most cases the CNLS method does not provide noise-immunity and it can be very sensitive to the experimental spectrum distortion provided as an input. 
So, to save high measurement rate conserving noise-immunity, we proposed the adaptive filtering EIS (AF-EIS) \cite{Stupin2017}.

\begin{figure}[h!]
    \centering
  \begin{tikzpicture}[ultra thick]

\draw  (-2.5,0) -- (0,1.25) ;
\draw [<-]   (0,1.25) -- (0,2.5)   node [right, align=center, midway]{$\times n_0$ } ;
\draw   (-2.5,0) -- (-1.2,1.25);
\draw [<-]   (-1.2,1.25) -- (-1.2,2.5) node [right, align=center, midway]{$\times n_1$ };
\draw    (-2.5,0) -- (-2.4,1.25);
\draw [<-]   (-2.4,1.25) -- (-2.4,2.5) node [right, align=center, midway]{$\times n_2$ };
\draw [dotted]  (-2.5,0) -- (-3.6,1.25);
\draw [<-,dotted]   (-3.6,1.25) -- (-3.6,2.5)node [right, align=center, midway]{...};
\draw   (-2.5,0) -- (-4.8,1.25);
\draw [<-]   (-4.8,1.25) -- (-4.8,2.5) node [right, align=center, midway]{$\times n_{\ell_n}$ };

\draw   (-2.5,0) -- (-1.2,-1.25);
\draw [<-]   (-1.2,-1.25) -- (-1.2,-2.5)node [right, align=center, midway]{$\times d_1$ };
\draw   (-2.5,0) -- (-2.4,-1.25);
\draw  [<-]  (-2.4,-1.25) -- (-2.4,-2.5) node [right, align=center, midway]{$\times d_2$} ;
\draw  [dotted] (-2.5,0) -- (-3.6,-1.25);
\draw  [<-,dotted]  (-3.6,-1.25) -- (-3.6,-2.5) node [right, align=center, midway]{...};
\draw   (-2.5,0) -- (-4.8,-1.25);
\draw [<-]   (-4.8,-1.25) -- (-4.8,-2.5)node [right, align=center, midway]{$\times d_{\ell_d}$ } ;
\draw [->]   (0,-2.5) -- (0,-0.35)node [right, align=center, midway]{$\times (-1)$ } ;

\draw [->]   (-2.5,0) -- (-0.85,0)node [above, align=center,very near end]{$J^a_k$ } ;

\draw [->, color=violet] (2,1.5).. controls (0.5,0.5) and (0.5,-0.5).. (2.5,-1.5);

\node at (0,2.5) [rectangle,draw=blue!50,fill=blue!20] {$V_{k\phantom{-1}}$};
\node at (-1.2,2.5) [rectangle,draw=blue!50,fill=blue!20] {$V_{k-1}$};
\node at (-2.4,2.5) [rectangle,draw=blue!50,fill=blue!20] {$V_{k-2}$};
\node at (-3.6,2.5) [rectangle,draw=blue!50,fill=blue!20] {\vphantom{$V_{k-2}$}...}; 
\node at (-4.8,2.5) [rectangle,draw=blue!50,fill=blue!20] {$V_{k-\ell_n}$};
\node at (1.2,2.5) [rectangle,draw=blue!50,fill=blue!20] {$V_{k+1}$};
\node at (2.4,2.5) [rectangle,draw=blue!50,fill=blue!20] {$V_{k+2}$};
\node at (3.6,2.5) [rectangle,draw=blue!50,fill=blue!20] {$V_{k+3}$};
\node at (4.8,2.5) [rectangle,draw=blue!50,fill=blue!20] {\vphantom{$V_{k-2}$}...}; 

\node at (0,-2.5) [rectangle,draw=red!50,fill=red!20] {$J_{k\phantom{-1}}$};
\node at (-1.2,-2.5) [rectangle,draw=red!50,fill=red!20] {$J_{k-1}$};
\node at (-2.4,-2.5) [rectangle,draw=red!50,fill=red!20] {$J_{k-2}$};
\node at (-3.6,-2.5) [rectangle,draw=red!50,fill=red!20] {\vphantom{$J_{k\phantom{-1}}$}...};
\node at (-4.8,-2.5) [rectangle,draw=red!50,fill=red!20] {$J_{k-\ell_d}$};
\node at (1.2,-2.5) [rectangle,draw=red!50,fill=red!20] {$J_{k+1}$};
\node at (2.4,-2.5) [rectangle,draw=red!50,fill=red!20] {$J_{k+2}$};
\node at (3.6,-2.5) [rectangle,draw=red!50,fill=red!20] {$J_{k+3}$};
\node at (4.8,-2.5) [rectangle,draw=red!50,fill=red!20] {\vphantom{$V_{k-2}$}...}; 

\node at (0,-4) [rectangle,draw=yellow!100,fill=yellow!50] {$\varepsilon_{k\phantom{-1}}$};
\node at (-1.2,-4) [rectangle,draw=yellow!100,fill=yellow!50] {$\varepsilon_{k-1}$};
\node at (-2.4,-4) [rectangle,draw=yellow!100,fill=yellow!50] {$\varepsilon_{k-2}$};
\node at (-3.6,-4) [rectangle,draw=yellow!100,fill=yellow!50] {\vphantom{$J_{k\phantom{-1}}$}...};
\node at (-4.8,-4) [rectangle,draw=yellow!100,fill=yellow!50]{$\varepsilon_{k-\ell_d}$};
\node at (1.2,-4) [rectangle,draw=yellow!100,fill=yellow!50] {$\varepsilon_{k+1}$};
\node at (2.4,-4) [rectangle,draw=yellow!100,fill=yellow!50] {$\varepsilon_{k+2}$};
\node at (3.6,-4) [rectangle,draw=yellow!100,fill=yellow!50] {$\varepsilon_{k+3}$};
\node at (4.8,-4) [rectangle,draw=yellow!100,fill=yellow!50] {\vphantom{$J_{k\phantom{-1}}$}...};
\node at (0,-3.3) [circle,draw=black!50,fill=white!50,top color=white, bottom color=black!20,minimum size=1 mm] {\tiny +};
\node at (-1.2,-3.3) [circle,draw=black!50,fill=white!50,top color=white, bottom color=black!20,minimum size=1 mm] {\tiny +};
\node at (-2.4,-3.3) [circle,draw=black!50,fill=white!50,top color=white, bottom color=black!20,minimum size=1 mm] {\tiny +};
\node at (-3.6,-3.3) [circle,draw=black!50,fill=white!50,top color=white, bottom color=black!20,minimum size=1 mm] {\tiny +};
\node at (-4.8,-3.3) [circle,draw=black!50,fill=white!50,top color=white, bottom color=black!20,minimum size=1 mm] {\tiny +};
\node at (1.2,-3.3)[circle,draw=black!50,fill=white!50,top color=white, bottom color=black!20,minimum size=1 mm] {\tiny +};
\node at (2.4,-3.3) [circle,draw=black!50,fill=white!50,top color=white, bottom color=black!20,minimum size=1 mm] {\tiny +};
\node at (3.6,-3.3) [circle,draw=black!50,fill=white!50,top color=white, bottom color=black!20,minimum size=1 mm] {\tiny +};
\node at (4.8,-3.3) [circle,draw=black!50,fill=white!50,top color=white, bottom color=black!20,minimum size=1 mm] {\tiny +};

\node at (-2.5,0) [circle,draw=green!50,fill=green!20] {$\sum$};

\node at (0.5,0) [rectangle,draw=red, fill=red!40] {$\delta_k=J^a_k-J_k$};

\node at (4,1.5) [rectangle,draw=green!50, fill=green!20] {\scriptsize{$J^a_k=\sum\limits^{\ell_n}_{j=0}n_j V_{k-j}+\sum\limits^{\ell_d}_{j=1}d_j J_{k-j}$}};
\node at (4,-1.5) [rectangle,draw=red!100, fill=red!75] {\small{$\sum\limits_{\mbox{\tiny by all $k$}}\delta_k^2=\min$}};
\end{tikzpicture} 
    \caption{The working principle of adaptive filtering EIS. At each step $k$, the adaptive filter takes several previous values of the excitation voltage and current responses of the unknown sample, multiplies them on the weight coefficients, sum result, and modifies weight coefficients in such a way to minimize the difference between adaptive filtering prediction $J^a_k$ and experimentally obtained current response $J_k$. After this procedure, the transfer function of the adaptive filter will be the noise-free approximation of the sample admittance. \label{Fig:AF_Scheme}}

\end{figure}
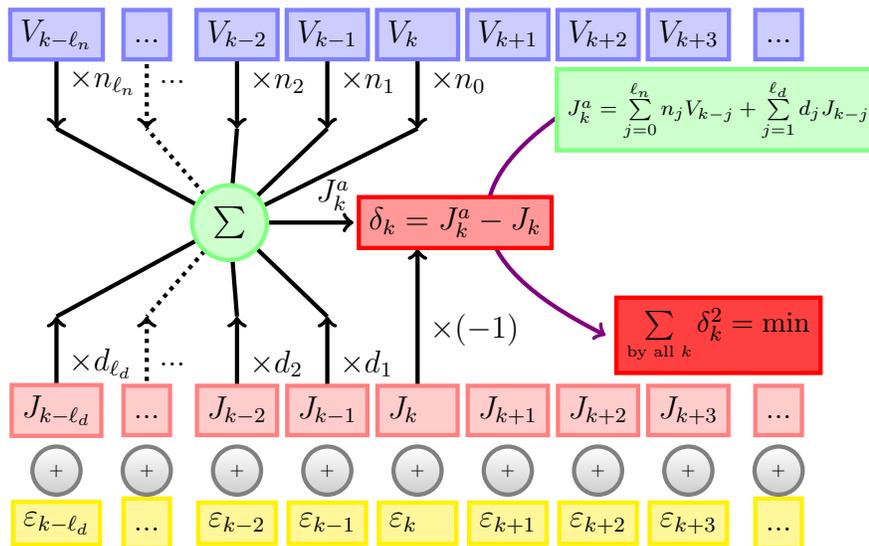

The key idea of the AF-EIS is that the current response and the excitation voltage are not independent, as in the case of Fourier-EIS, and are related to each other via a linear equation in finite differences. Or in other words, the current response is formed by processing excitation voltage with a digital filter.
So, if we guess what filter converts the excitation voltage into the current response, we can easily find its transfer function, which is obviously  equal to  admittance of the sample. 
This problem is known as the problem of a linear system identification or a black-box identification.
A powerful tool for solving this problem is the adaptive filtering technique \cite{Stearns, widrow1975adaptive}.
From the mathematical point of view this problem can be formulated as follows (Fig. \ref{Fig:AF_Scheme}). 
Given the $V_k$ and $J_k$ sequences, and we should construct such filter 

\begin{equation}
    J^a_k=\sum\limits^{\ell_n}_{j=0}n_jV_{k-j}+\sum\limits^{\ell_d}_{j=1}d_j J_{k-j},
\end{equation}
that minimises the functional
\begin{equation}
\label{Eq:minimisation}
    \sum\limits_{\mbox{\tiny  by all $k$}}|J_k-J^a_k|^2=\min.
\end{equation}

Here, $J_k^a$ is the prediction of the adaptive filter, $n_j$ and $d_j$ are so-called weight coefficients, and $\max(\ell_n, \ell_d)$ is the filter order.
 When optimal weight coefficients are found (after filter learning process), with a high enough sampling rate and high enough data collection time, the admittance of the sample can be simply calculated with the formula 
\begin{equation}
\label{Eq:AF_Y}
    Y \approx\left(\sum_{j=0}^{\ell_n} n_j \varphi_j \right) \left/ \left( 1-\sum\limits_{j=1}^{\ell_d} d_j \varphi_j\right) \right.,
\end{equation}
where $\varphi_j=\exp{[i2 \pi (f/f_0) j]}$, $f$ is the frequency, and $f_0$ is the sampling rate. 
In fact, Eq.~\ref{Eq:AF_Y} is the Levy \cite{Levy} approximation of the sample admittance in the $[-f_B, f_B]$ range ($f_B$ is the highest harmonic of the excitation voltage) with its weight equals to the square of the excitation voltage spectrum magnitude.
The main advantage of the AF-EIS compared to Fourier-EIS is the high noise immunity.
It can be shown that in case of auto-correlated noise that is not correlated with the excitation voltage and the current response, the error in weight coefficients and, as a consequence, the error in the obtained immittance spectrum is

\begin{equation}
    \mbox{Error in WC} \leq C\frac{\langle \varepsilon^2 \rangle }{\sqrt{\ell_n+\ell_d+1}}, 
\end{equation}
where $\langle \varepsilon^2 \rangle$ is noise mean square value, and $C$ is a multiplayer, which value depends on the conditioning of the problem \eqref{Eq:minimisation} \cite{Stupin2017}. The AF-EIS could be applicable for single living cell research even in high-noise environment (signal-to-noise ratio 3 dB) \cite{stupin2018single,Stupin2017,stupin2017single, stupin2018tin}.

\section{EIS applications in biology and medicine}
In this section, we will briefly describe several applications of EIS in biology, medicine, pharmacology, and biosensorics.

\subsection{Living cells research}
One of the most popular EIS biological applications  is the research on living cells. 
These studies are usually based on the ECIS-type electrodes geometry, where cells are grown on the planar electrodes located on the bottom of the modified Petri dish (see Sec. \ref{Sec:Giaever-Keese_model}).
The analysis of time-resolved electrodes  immittance response  allows the researchers to estimate the number of cells on the electrode, the motility or viability of cells, \textit{etc}. 
Up to date, two commercially-available systems  that can be used to perform this tasks exist: ECIS (Applied Biophysics, USA) and xCelligence (ACEA Bioscience, USA). Further we will deal with last one device.

\begin{figure}
    \centering
    \includegraphics[scale=0.5]{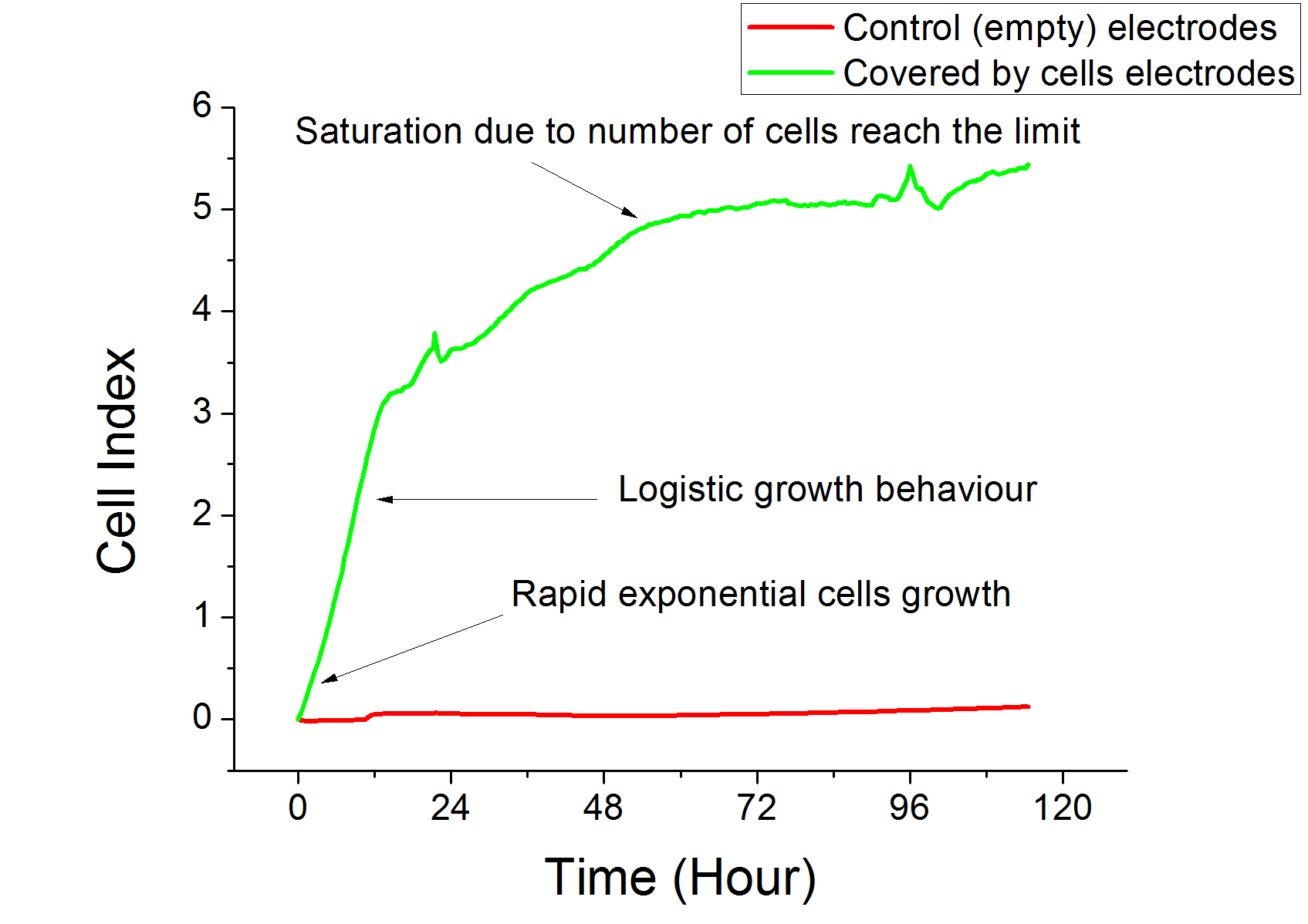}
    \caption{The impedance evolution of a cell-covered electrode (green) and an empty electrode (red). During fission and growth of cells the impedance of the cell-covered electrode demonstrates a logistic growth. In contrast the impedance of the empty electrode is stable in time. This result confirm Giaever-Keese model (Sec. \ref{Sec:Giaever-Keese_model}). \label{Fig:xCelligence_grown}}

\end{figure}

Despite of the xCelligence in fact produces immittance measurement at the only one frequency (10 kHz, \cite{peters2015human}), and thus it  generally speaking it is not spectroscope,  it can be successfully  used in number of the actual cells-research applications.
Here we should emphasise that  as an output result xCelligence provides a normalized impedance that is called the \textit{Cell Index}

\begin{equation}
    \mbox{Cell Index}=\frac{Z(t)-Z(0)}{Z(0)},
\end{equation}
where $Z(t)$ is the magnitude of the impedance at frequency 10 kHz at time $t$ and $Z(0)$ is the magnitude of the impedance at the beginning of the experiment.
One of  possible xCelligence application -- building  growth curve of the HeLa cells \cite{masters2002hela} -- is shown on the Fig.~\ref{Fig:xCelligence_grown}. It depicts the impedance evolutions of two electrodes: one is covered with cells and another is the  control one (cells-free).
One can see that the Cell Index of the empty electrode is stable in time, however, the Cell Index of the cell-covered electrode demonstrates a logistic rise \cite{escudero2004extinction}, which is related to cell fission and growth.
So, the data presented in Fig. \ref{Fig:xCelligence_grown}  qualitatively confirms the Giaever-Keese model. 

\begin{figure}
    \centering
    \includegraphics[scale=0.5]{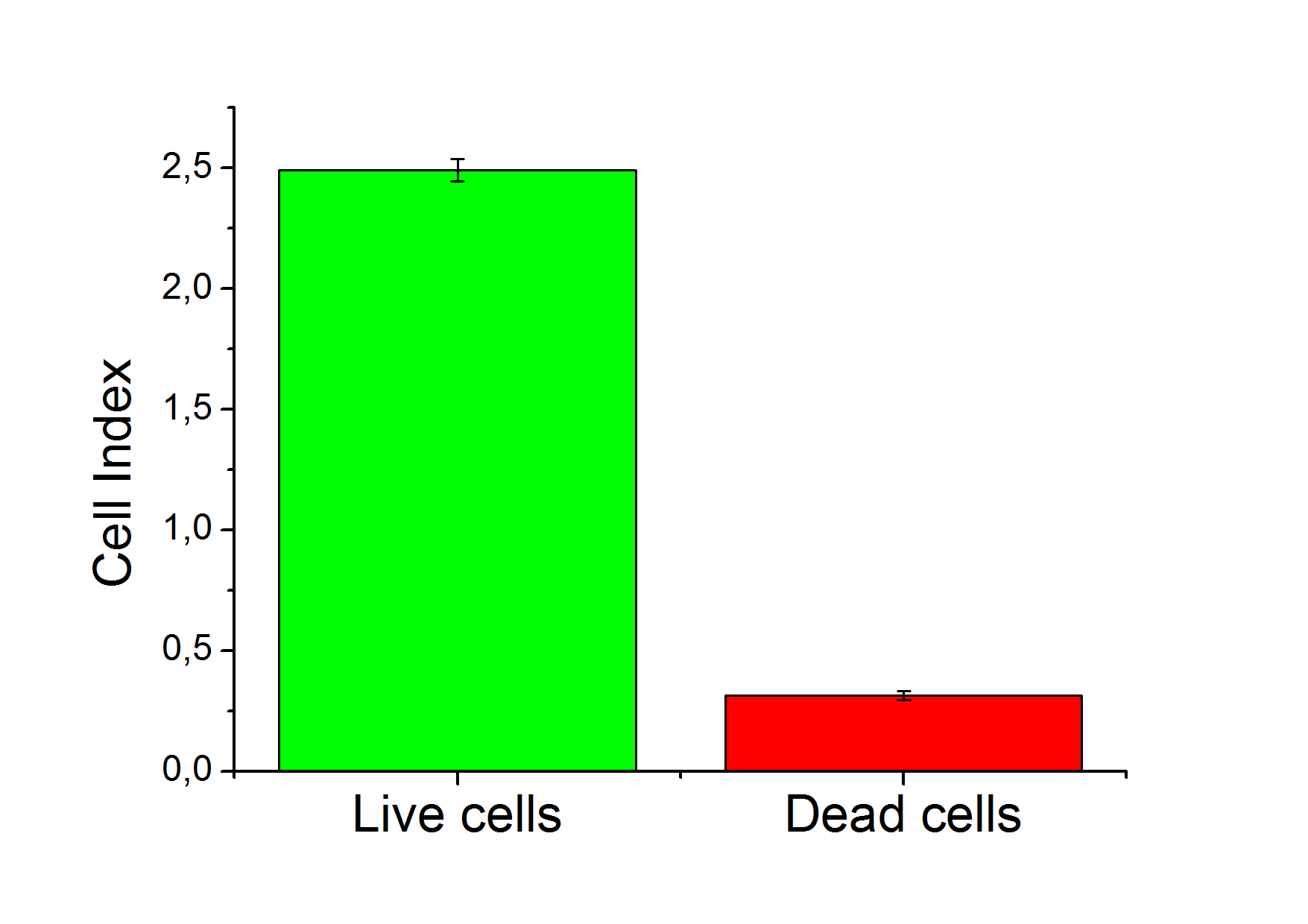}
    \caption{The effect of a toxin addition on the metal/electrolyte/cell impedance measured with xCelligence at frequency 10 kHz. One can see, that the electrode covered by dead cells has significantly lower impedance compared to the impedance of the electrode with alive cells. Data was averaged over three measurements, the error bars correspond to 99.9\% confidence interval.\label{Fig:xCelligence_toxin}}
   \end{figure}

\begin{figure}
    \centering
    \subfigure[]{\includegraphics[scale=0.5]{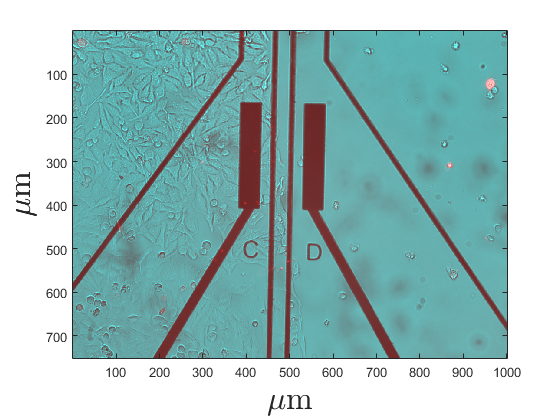}}\subfigure[]{\includegraphics[scale=0.5]{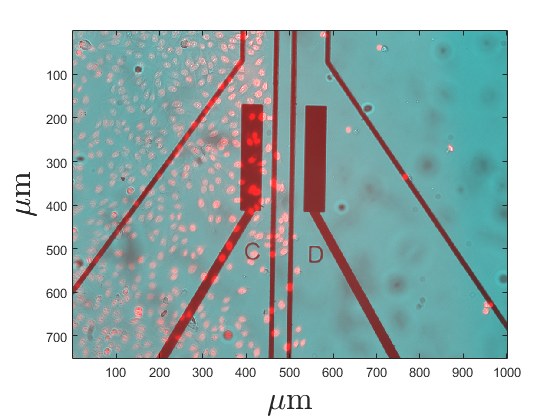}}
    \subfigure[]{\includegraphics[scale=0.5]{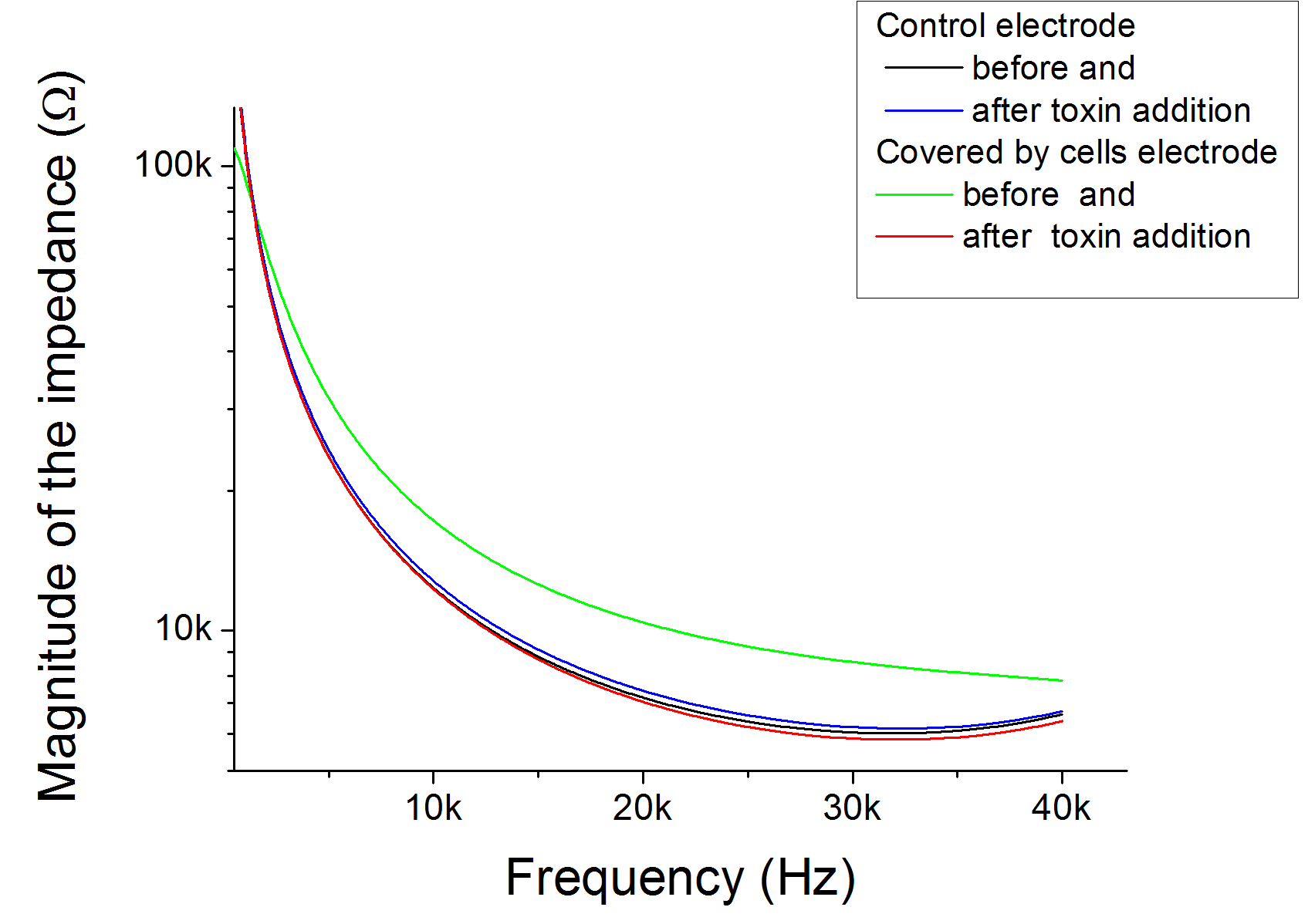}}
    \caption{The effect of a toxin addition on the metal/electrolyte/cell interface measured in broad frequency range. (a) and (b) a pseudo-photograph of cells before  and after  the addition of the toxin respectively. The red channel corresponds to propidium iodide dye \cite{bank1988rapid} fluorescence in the nuclei of dead cells. Blue channel corresponds to the bright field image in transfer light. (c) Obtained impedance spectra. One can see that after the addition of a toxin the high-frequency part of the impedance of the cell-covered electrode decreases to the level of the empty control electrode. However, the low-frequency part of the spectrum demonstrates a capacitance-like behavior, which means that electrodes are not destroyed, and the impedance decrease is indeed caused by biological effects. Spectra are averaged over three measurements. The obtained confidence intervals for 99.9\% reliability is very small, and thus omitted from the plot.}\label{Fig:AF_toxin} 
\end{figure}

Another important bio-EIS application for cell research is the identification of the cell state.
As an example, we will describe an experiment that was performed both with the xCelligence device and with a home-built device based on the multielectrode array 60StimMEA200/30-Ti (Multichannel Systems, Germany) and the adaptive filtering for signal analysis \cite{Stupin2017}. 
As a sample we again have used the HeLa cancer cells. 
The experimental protocol was trivial: we measured the immittance of the metal/electrolyte/cell interface before and after the addition of the toxin Triton x100 (Union Carbide, USA) to cell medium. The results are presented on Figs. \ref{Fig:xCelligence_toxin} and \ref{Fig:AF_toxin}.
As one can see  from Fig. \ref{Fig:xCelligence_toxin} the Cell Indexes measured by xCelligence before and after the addition of the toxin are  distinctly different. This effect is confirms Giaever-Keese model: due to membrane distortion of dead cells the electrode covered with them has lower impedance compared to the electrode covered with alive cells (see Fig. \ref{Fig:Giaever_Keese}). Thus the EIS allows for distinguishing between dead and alive cells.
However, xCelligence provides measurements only at a single frequency, and for this reason, it does not provide a reliable cell viability estimation.
\textit{I.e.,} a picture similar to the  Fig. \ref{Fig:xCelligence_toxin} can be obtained by destroying the dielectric coating of the electrodes or by increasing the conductivity of cell medium after a toxin addition.  
To significantly increase the reliability of EIS as a method for estimating cell viability, one has to perform measurements in a broad frequency range and to use the control (empty) electrode that is located in the same medium \cite{stupin2018cell}. 
These two modifications will allow the researcher to distinguish biological and physical/chemical phenomena affecting immittance. 

The clear advantages of the broad-frequency range approach with the usage of multiple electrodes are shown in Fig. \ref{Fig:AF_toxin}. 
One can see that the impedance of the control electrode is stable upon the addition of a toxin, thus the conductivity of the medium does not changed during the experiment. 
On the contrary, for the electrode with cells, the high-frequency part of the impedance decreases upon the addition of the toxin. 
This drop of impedance is directly caused by the destruction of the cell membrane, \textit{i.e.} cell death [see Fig. \ref{Fig:Giaever_Keese}(a)], which is also confirmed by fluorescence microscopy data [Fig.~\ref{Fig:AF_toxin}(a,~b)].
In the same time, the low-frequency part of the electrodes' spectra demonstrates a capacitance-like behaviour [see Eqs. \eqref{Eq:GK-model} and \eqref{Eq:Z_Dispersion}], so the electrodes does not destruct during the experiment. 
Thus, compared to single-sine EIS methods, a broad-frequency EIS approach based on a multielectrode array provides reliable data that is robust to non-biological phenomena.


\subsection{Medical applications}

EIS also finds a wide range of applications in medicine and healthcare.
One of the examples is \textit{impedance cardiography} -- the monitoring of heart activity by measuring the impedance of a patient's thorax.
In literature \cite{ICG} two mechanisms that affect impedance variation \textit{v.s.} cardiac systole are considered. 
The first mechanism is based on the fact that blood has high conductivity compared to tissues. 
Thus, the increase of blood volume in thorax results in a decrease in its resistance. 
The second mechanism involves the phenomenon of erythrocytes orientation.
Due to mechanical stress in the blood, the disk-shaped erythrocytes arrange parallel to the blood flux, what results in decreasing blood, and thus thorax, resistance. 
Therefore, impedance cardiography reflects the pumping activity of the heart, contrary to electrocardiography that reflects the heart's electrical activity.
An obvious advantage of impedance cardiography compared to electrocardiography is a high noise immunity. 
Impedance cardiography signal can be increased to higher amplitudes relative to electrocardiography amplitudes and it is robust to the interference with industrial frequencies (50 or 60 Hz) because impedance cardiography measurement is performed at higher frequencies.

Another medical application of the EIS is \textit{pathology diagnostics}, including cancer detection. 
Even though the tissue impedance spectrum can vary for different patients, for disease detection one can compare the impedance spectra of paired organs, one of which is healthy, and the second one is pathological. For example, typically cancer tissue has lower resistance relative to normal tissue.
This phenomena and based on it technique is applied for the early detection of breast cancer \cite{zou2003review}. The difference in the impedances of the pathology and healthy tissues is also used for skin cancer detection. These diagnostic techniques are performed by using multielectrode sensor for \textit{in vivo} impedance measurement of the two skin regions -- the region of interest, which possibly contain tumor, and healthy neighborhood region. For example, if cancer test provided on the arm's skin it is possible to use contralateral regions of it for such probing. As reported in Refs. \cite{braun2017electrical,moqadam2018cancer} impedance-based skin cancer detection allows to distinguish healthy and pathology skin regions, which open a new perspectives in oncology monitoring. It also should be notices, that EIS is widelly use not only in cancer diagnostics, but also in cancer research \cite{Giaever_book}. 

    The fact that different tissues have different conductivity could be used for providing \textit{impedance tomography}, \textit{i.e.} for visualizing  in-body space by impedance measurements \cite{brown2001medical}. This diagnostic technique usually utilizes  an array of electrodes that are placed on the body around the volume of interest, for example around the thorax. The impedances of electrodes are measured and then by applying special reconstructing algorithms \cite{harikumar2013electrical} their values are used for building the 3D-image of the body region that is being examined. This type of tomography  provide real-time, inexpensive, and easy in implementation diagnostics. Its another advantage is being hazard-free, because low currents are safe for human unlike, for example, x-rays. This non-hazard property of the impedance tomography is extremely useful for monitoring health of the patients with pulmonology as well as health of the attending them pulmonologists.
    Moreover, with the usage of multi frequency impedance tomography it is possible to provide additional tissues characterization as it was mentioned above. For instance, impedance tomography  applicable for detection breast cancer even without comparison with health gland \cite{kerner2002electrical}. The another applications of the impedance tomography involved brain visualization, temperature treatment control in tumor  oncotherapy,  gastrointestinal tract diagnostics, and thorax monitoring in aircraft industry  \cite{bayford2006bioimpedance}.

Impedance spectroscopy is also used in \textit{electronic implants} as a label-free characterization tool. 
For example, impedance measurements are used in vision prosthetic care Argus II device for the diagnostics of the interface between retina and stimulus electrodes \cite{luo2013mri}.
This interface defines the quality of the patient's visual perception and it could be also described by Giaever-Keese model. 
If the electrodes are in good contact with retina cells, their impedance magnitude will be relatively high. 
On the contrary, if electrodes have a low impedance magnitude, it means that the implant stimulus array is detached from the retina, and the patient should consult with a doctor. 
Finally, if the electrode shows a very high impedance magnitude, it means that the electrode is burnt-out or damaged. 
In this case, Argus II implant electronics will switch off the failure electrode and switch on one of neighbouring reserved electrodes. 
It should be stressed out that in vision and hearing prosthetic care the impedance of the stimulus electrodes is their main characteristic, and engineers try to make it as small as possible \cite{franks2005impedance}.

\subsection{Bio-matter and biosensorics}
As a non-optical, label-free, sensitive, and easy in implementation technique, EIS found many applications in biosensorics and biological matter studies. 
For example, EIS was proposed as a method for the \textit{detection of glucose} in biological samples.
In the work  \cite{tlili2003fibroblast}, the ECIS-like device allowed the authors to detect glucose with 3T3-L1 fibroblast cells.
Another approach was proposed in the work \cite{pradhan2019quantitative}. 
The authors showed a quantitative difference in impedance spectra of the metal/blood interface for different glucose concentrations in blood samples. 
This technique can be useful for portable on-body devices like system reported in \cite{liao20113}. 

EIS is also used for the investigation of \textit{biomolecules} \cite{hou2007novel, cornell2012comparative, li2007kinetic, long2003ac}. 
Particularly, in Ref. \cite{hou2007novel} and \cite{alfinito2010role} the metal/protein interface was thoroughly investigated. 
It was demonstrated that the immittance of the metal/protein interface significantly depends on the protein structure and state.
Along with the impedance network protein analog approach \cite{alfinito2010role,alfinito2008network}, these results may pave the road for the development of new EIS based tools for protein investigation.

One more promising EIS bio-sensing application is \textit{virus detection}. Several scientific groups recently have shown, that it is possible to detect viruses by measuring the impedance of the electrodes, which surface is modified by antibodies \cite{nidzworski2014universal,diouani2008miniaturized,hassen2011quantitation}. These sensor approaches relatively inexpensive, easy-in-implementation, and could be realized in portable devices, which is important for personalized medicine. The authors also reported that EIS provides sensitive (detection limits 8 ng/ml \cite{hassen2011quantitation} and $\sim 100$ virions in $\mu$l \cite{nidzworski2014universal}) and fast (in $\sim 30$ min \cite{nidzworski2014universal}) virus detection in near to physiological envelopment. Additionally, proposed in Ref. \cite{diouani2008miniaturized} phenomenological EC for anti-body modified electrodes together with CNLS  fitting [Eq. \eqref{Eq:CNLS}] open way for more informative and reliable virus biosensing technologies. Another scientific groups \cite{campbell2007monitoring,mccoy2005use,pennington2017electric,cho2007impedance,golke2012xcelligence,teng2013real} propose to use cell-based EIS virus sensors, which working principle is directly grounded on the Giaever-Keese model [see Fig. \ref{Fig:Giaever_Keese}]: the addition of the viruses to the cells' medium results in cell infection and death what dramatically reduce the impedance of covered by them electrodes. Thus, by monitoring electrode/cell interface impedance it is possible to detect viruses without special electrode preparation with antibodies and other surface modification reagents. 
Moreover, cell-based EIS virus sensors in contrast to convenient ELISA kits and real-time PCR methods \cite{Immunology_book,shojaei2015review} can react on the unknown types of viruses and also could be utilized several times, because the their  virus-detection properties could be repaired by treatment electrodes  in high-pressure steam sterilizer with following cells seeding on them.
It also should be stressed, that nowadays cell-based biosensors  found several applications in virus disease research \cite{de2017serum,al2017japanese,jiang2010electrochemical}. 
The information about completely  label-free EIS virus detectors is presented in Refs. \cite{poenar2014label,hatsuki2015nonlinear}.  To summarize, the EIS-based virus sensing technologies provide a hopeful tool for fighting with various virus diseases like immunodeficiency syndrome, dengue virus-infection,  and, highly likely,  COVID19.

Finally, it is worth mentioning that EIS can be used to study \textit{cell membranes} in combination with patch-clamp technology \cite{Fuchs}. 
In this diagnostic technique, a cell membrane is attached to the glass microtube with a diameter of $\sim\mu$m that is filled with electrolyte. 
Two AgCl electrodes are placed in the microtube and the cell medium, and the impedance is measured between these electrodes. 
Because the impedance of the membrane is significantly higher than the impedance of the electrodes/electrolyte interface (Tab. \ref{tab:Cell_electrical_properties}), its value will define the immittance between the electrodes. 
As a result, the combination of a patch-clamp with EIS allows the researchers to evaluate the conductivity of the cell membrane, the resistivity of ion channels in cell membrane and other important electrical characteristics of a cell.
For example, in Ref. \cite{matsumura2018dependence} this technique based on Fourier-EIS allowed the authors to detect that living neurons can regulate their capacitance in response to changes of membrane area.

\section{Conclusion}
In this tutorial article, we reviewed modern electrical impedance spectroscopy  from the theoretical and experimental points of view, paying special attention to its applications in cell research, medicine, including viruses and cancer study, and biosensing. 
We hope that this article will be useful  for students   as rapid introduction  to EIS as well as for scientists and engineers as EIS handbook.
Nevertheless, we do  not expect that our study can completely replace classical and modern works on EIS.
For further reading, the following papers can be used. 
The books by E. Barsoukov and J. Ross Macdonald \cite{Barsoukov} and V.F. Lvovich \cite{Lvovich} thoroughly describe electrical impedance spectroscopy and its application for electrochemistry. 
The book by S. Grimnes and O.G. Martinsen contains a lot of information about bio-impedance \cite{BioImp}. 
The review \cite{Review} by B.Y. Chang and S.M. Park is devoted to modern high-speed EIS technique based on Fourier transformation.
The description of another high-speed technique, adaptive filtering EIS, can be found in the article \cite{Stupin2017}.
The article \cite{MacdonaldR} by J. Ross Macdonald is devoted to complex non-linear least-squares approximation -- a state-of-art tool for EIS spectra analysis.
Chapters 3.4 and 7 in Ref. \cite{Cell_based_biosensors} by P. Wang and Q. Liu contain information on biosensorics based on EIS. 
The book \cite{Giaever_book} by Wen G. Jiang contains information about cellular biology research conducted with EIS. The review by R.H. Bayford is concerned with the progress in impedance tomography. 
Titche and Shnek book \cite{Tietze} introduces the reader into EIS basics and its measuring equipment.
The most simple and elegant introduction into EIS can be found in Richard Feynman's lectures \cite{Feynman_Lectures} in chapter 25-5.



\ack
The authors acknowledge  N.A. Verlov, N.A. Knyazev, A.A. Kornev, and A.V. Naschekin for comprehensive  assistance and support.

This work was supported by the Ministry of Education and Science of Russian Federation.
The comparative research of the single-frequency and single-electrode biosensors \textit{v.s.} multiple-frequency and multielectrode biosensors was funded by RFBR according to the research project  18-32-00363.

D.M.N. and M.N.R. acknowledge HPC computing resources at the Lomonosov Moscow State University and Resource Center ''Computer Center of SPbU''. 
\section*{References}



\end{document}